\documentclass[preprintnumbers, floatfix, preprintnumbers, letterpaper, superscriptaddress,nofootinbib, onecolumn]{revtex4}
\pdfoutput=1
\usepackage{graphicx}
\usepackage{amsmath}
\usepackage{amssymb}
\usepackage{subfigure}
\usepackage{hyperref}
\usepackage{url}
\usepackage{xcolor}
\usepackage{color}
\usepackage{mathrsfs}
\usepackage{calrsfs}
\usepackage{amsfonts}
\usepackage{latexsym}
\usepackage{ragged2e}
\usepackage{textcomp}
\usepackage{phaistos}
\usepackage{epstopdf}

\usepackage{soul,color}
\soulregister\cite7
\soulregister\ref7
\soulregister\pageref7

\makeatletter
\renewcommand\@makefnmark{\hbox{\@textsuperscript{\normalfont\color{purple}\@thefnmark}}}
\renewcommand\@makefntext[1]{%
  \parindent 1em\noindent
            \hb@xt@1.8em{%
                \hss\@textsuperscript{\normalfont\@thefnmark}}#1}
\makeatother
\usepackage{caption}
\DeclareCaptionJustification{justified}{\leftskip=0pt \rightskip=0pt \parfillskip=0pt plus 1fil}
\definecolor{vividviolet}{rgb}{0.62, 0.0, 1.0}
\definecolor{amaranth}{rgb}{0.9, 0.17, 0.31}
\definecolor{palatinateblue}{rgb}{0.15, 0.23, 0.89}
\definecolor{brightpink}{rgb}{1.0, 0.0, 0.5}
\definecolor{cornflowerblue}{rgb}{0.39, 0.58, 0.93}
\definecolor{deepcarminepink}{rgb}{0.94, 0.19, 0.22}
\definecolor{radicalred}{rgb}{1.0, 0.21, 0.37}

\hypersetup{ linktoc=all,
    colorlinks, linkcolor={palatinateblue},
    citecolor={brightpink}, urlcolor={amaranth}
}

\graphicspath{{Images/}}

\graphicspath{{Images/}}

\makeatletter
\def\@fnsymbol#1{\ensuremath{\ifcase#1\or $\PHplaneTree$ \or $\textleaf$ 
\else\@ctrerr\fi}}%

\makeatother
\def\sideremark#1{\ifvmode\leavevmode\fi\vadjust{\vbox to0pt{\vss% the remark
 \hbox to 0pt{\hskip\hsize\hskip1em%                          will appear only
 \vbox{\hsize1.5cm\tiny\raggedright\pretolerance10000%          on the side
 \noindent #1\hfill}\hss}\vbox to8pt{\vfil}\vss}}}%
\def\sideremark#1{\ifvmode\leavevmode\fi\vadjust{\vbox to0pt{\vss% the remark
 \hbox to 0pt{\hskip\hsize\hskip1em%                          will appear only
 \vbox{\hsize1.3cm\tiny\raggedright\pretolerance10000%          on the side
 \noindent #1\hfill}\hss}\vbox to8pt{\vfil}\vss}}}%
\begin{document}

%\title{Wormhole Immersed in the Magnetic Monopole (?? OK or not? Make Sense?)}
\title{Non--Abelian Wormholes Threaded by a Yang-Mills-Higgs Field in the BPS Limit}

\author{Xiao Yan \surname{Chew}}
\email{xychew998@gmail.com, xiao.yan.chew@uni-oldenburg.de}
\affiliation{Department of Physics Education, Pusan National University, Busan 46241, Republic of Korea}
\affiliation{Research Center for Dielectric and Advanced Matter Physics, Pusan National University, Busan 46241, Republic of Korea}

\author{Kok-Geng \surname{Lim}}
\email{K.G.Lim@soton.ac.uk}
\affiliation{University of Southampton Malaysia, 79200 Iskandar Puteri, Johor, Malaysia}
%\affiliation{School of Aeronautics and Astronautics, Shanghai Jiao Tong University, Shanghai 200240, China}
%\affiliation{Nordita, KTH Royal Institute of Technology \& Stockholm University,
%Roslagstullsbacken 23, SE-106 91 Stockholm, Sweden}

\begin{abstract}
In the Bogomol'nyi--Prasad--Sommerfield (BPS) limit of Einstein--Yang--Mills--Higgs theory, we construct numerically a non--Abelian wormhole supported by a phantom field. The probe limit is the Yang--Mills--Higgs field in the background of Ellis wormhole when the gravity is switched off. The wormhole solutions possess the Yang--Mills--Higgs hair in the presence of gravity; Thus a branch of hairy wormhole solutions emerge from the Ellis wormhole when the gravitational coupling constant increases. We find that the mass of wormholes and scalar charge of phantom field increase monotonically when the gravitational coupling constant increases. The wormhole spacetime possesses double throat configuration when the gravitational strength exceeds a critical value. Surprisingly, the wormholes satisfy the null energy condition in the large gravitational strength. We also briefly discuss the redshift factor.
\begin{center}

\end{center}
\end{abstract}

\maketitle

\section{Introduction}

It is well known that the SU(2) Yang--Mills--Higgs (YMH) theory with Higgs field in the adjoint representation possesses both the magnetic monopole  \cite{t1974magnetic,polyakov1974particle} and multimonopole \cite{rebbi1980multimonopole,ward1981yang,forgacs1981exact,prasad1981construction} solutions with finite energy. The 't Hooft--Polyakov magnetic monopole \cite{t1974magnetic,polyakov1974particle} is the first solitonic monopole solution in non--Abelian YMH theory with spontaneously broken symmetry. The exact solutions of magnetic monopole and multimonopole are only found in the Bogomol'nyi--Prasad--Sommerfield (BPS) limit with vanishing Higgs self--interaction potential and satisfy the first order Bogomol'nyi equation \cite{prasad1975exact,bogomol1976stability}. Their mass also saturates the lower bound, which is the Bogomol'nyi bound \cite{bogomol1976stability}. However, the numerical solutions could only be found beyond the BPS limit when the Higgs potential is non--vanishing \cite{kleihaus1998interaction}. There exist unstable and saddle point solutions which do not satisfy the Bogomol'nyi equation, but they are only solutions to the second order equations of motion of YMH theory. For instance, the monopole--antimonopole pair \cite{kleihaus1999monopole} and monopole--antimonopole chain solutions \cite{kleihaus2003monopole,kleihaus2004monopole}.

In the Einstein--Yang--Millls--Higgs (EYMH) system where the YMH field coupled with gravity, a branch of gravitating monopole solutions emerges from 't Hooft--Polyakov monopole in the flat space \cite{Breitenlohner:1991aa,lee1992black,Breitenlohner:1994di,Brihaye:1998cm,Lue:1999zp,Brihaye:1999nn,Brihaye:1999kt,
Brihaye:1999zt,Brihaye:2002pc}. In the BPS limit, the mass of the gravitating monopole decreases monotonically as the gravitational strength increases. When the gravitational strength reaches to a critical value, the solutions of gravitating monopole end up as an extremal Reissner--Nordstrom black hole. This also occurs for the static axially symmetric gravitating monopole solutions \cite{Hartmann:2000gx,Hartmann:2001ic}. The counterpart EYMH black holes also exist, and they possess the non--abelian gauge field outside the event horizon. Hence, they are also dubbed as the ``black hole within monopole'' \cite{lee1992black,Brihaye:1998cm}, which is a counterexample to the ``no hair'' conjecture for black holes. The monopole--antimonopole pairs \cite{Kleihaus:2000hx,Kleihaus:2000kv,Hartmann:2001ic,VanderBij:2001nm,Paturyan:2003hz,Paturyan:2004ps,Kleihaus:2004gm} and vortex rings \cite{Kleihaus:2004fh,Kleihaus:2005fs} also can be constructed in EYMH system.

Recently, a type of non--abelian wormhole has been obtained numerically in Einstein--Yang--Mills (EYM) system where the throat of wormhole is supported by a phantom field \cite{hauser2014hairy}. These hairy wormholes solutions possess a sequences of solutions, which are labelled by the node number $k$ of the gauge field function. They are analogous to the Bartnik--McKinnon solution, which is the regular and spherically symmetric solutions of the EYM system \cite{bartnik1988particlelike}. A phantom field is a real--valued scalar field which has an opposite sign for the kinetic term. It can be used to model the accelerated expansion of our universe in the cosmology \cite{Caldwell:1999ew,Carroll:2003st,Gibbons:2003yj,Hannestad:2005fg} and construct some compact objects such as  black holes \cite{Bronnikov:2005gm,Chen:2016yey}, black rings \cite{Kleihaus:2019wck}, star--like objects \cite{Dzhunushaliev:2008bq} and wormholes \cite{Ellis:1973yv,Ellis:1979bh,Bronnikov:1973fh}.

The construction of traversable wormholes in GR usually requires the violation of energy condition \cite{Visser:1995cc} to prevent the collapse of the throat \footnote{However, there are wormhole solutions can be constructed in the modified theory of gravity without the exotic matter \cite{Kanti:2011jz,Kanti:2011yv,Antoniou:2019awm}}. A classic example of such traversable wormhole is the static Ellis wormhole, which is supported by the phantom field \cite{Ellis:1973yv,Ellis:1979bh,Bronnikov:1973fh}. However, the Ellis wormhole possesses unstable radial modes \cite{Gonzalez:2008wd,Gonzalez:2008xk,Torii:2013xba}. The static Ellis wormhole has also been generalized to the higher--dimensional case \cite{Torii:2013xba}, the slowly rotating case with perturbative method \cite{Kashargin:2007mm,Kashargin:2008pk}, the rapidly rotating case in four dimensions \cite{Kleihaus:2014dla,Chew:2016epf} and five dimensions with equal angular momenta \cite{Dzhunushaliev:2013jja}, and also in the modified gravity, e.g., the scalar--tensor theory \cite{Chew:2018vjp}.

Furthermore, several astrophysical signatures of wormholes have been proposed to search for their existence in near future, since they might mimic black holes, for example, the shadow \cite{Nedkova:2013msa,Gyulchev:2018fmd,Amir:2018szm}, the gravitational lensing \cite{Abe:2010ap,Toki:2011zu,Takahashi:2013jqa,Cramer:1994qj,Perlick:2003vg,Tsukamoto:2012xs,Bambi:2013nla}, the accretion disk around the wormhole \cite{Zhou:2016koy}, and the ringdown phase in the emission of gravitational waves \cite{Blazquez-Salcedo:2018ipc}.

Although there are also some wormhole--like structures in EYMH theory reported in \cite{Hajicek_1983a,Hajicek_1983,1987GReGr..19..739D}, the wormholes in EYMH theory with the phantom field has not been explored yet. In this paper, our motivation is based on the solutions of particle--like and black holes in EYMH theory give rise to new and interesting phenomena due to the presence of non--Abelian field. Since the counterpart EYMH black hole exists, then the wormhole configuration with non--Abelian YMH hair should also exist. Thus, we follow the approach of \cite{hauser2014hairy} to numerically obtain the symmetric wormhole solutions in EYMH with a phantom field in the BPS limit and study their properties in this paper. Our paper is organized as follows. In section II, we briefly introduce the YMH theory and present the equation of motions. Subsequently, we derive the ordinary differential equations (ODEs) from the equation of motions. We then introduce the geometrical properties, the global charges, the null energy condition of wormholes and the boundary conditions imposed to the ODEs. In section III, we exhibit and discuss our numerical results. In section IV, we conclude our research works and briefly discuss the stability of the solutions and the possible outlook from this present work.

\section{Theoretical Framework}

\subsection{Theory and Ans\"atze}

In the Einstein--Hilbert action, we consider Einstein gravity to be coupled with a phantom field $\psi$ and a gauge field $A_\mu$ in SU(2) YMH theory in BPS limit,
\begin{equation} \label{EHaction}
 S_{\text{EH}}=  \int d^4 x \sqrt{-g}  \left[  \frac{R}{16 \pi G}  + \mathcal{L}_{\text{ph}} + \mathcal{L}_{\text{YMH}} \right]  \,,
\end{equation}
where the Lagrangian of phantom field and YMH \cite{Hartmann:2001ic} are respectively, given by 
\begin{equation}
  \mathcal{L}_{\text{ph}} = \frac{1}{2} \partial_\mu \psi \partial^\mu \psi \,, \quad \mathcal{L}_{\text{YMH}} =  - \frac{1}{2} \text{Tr}\left( F_{\mu \nu} F^{\mu \nu} \right) - \frac{1}{4} \text{Tr} \left(  D_{\mu} \Phi D^{\mu} \Phi  \right)   \,.
\end{equation}
The covariant derivative of the Higgs field and the gauge field strength tensor are given respectively by 
\begin{align}
F_{\mu\nu} &= \partial_{\mu}A_\nu - \partial_{\nu}A_\mu + i e \left[ A_{\mu}, A_\nu \right] \,, \\
D_{\mu} \Phi &= \partial_{\mu} \Phi + i e \left[ A_{\mu}, \Phi \right] \,, 
\label{1.1.2}
\end{align}
where $A_\mu = \frac{1}{2} \tau^a A_\mu^a$ and $\Phi=\phi^a \tau^a$ with $\tau^a$ is the Pauli matrices.

By varying the action with respect to the metric $g_{\mu \nu}$, we obtain the Einstein equation,
\begin{equation} \label{einstein_eqn}
 R_{\mu \nu} - \frac{1}{2} g_{\mu \nu} R =  \beta \left( T_{\mu \nu}^{\text{ph}} + T_{\mu \nu}^{\text{YMH}} \right) \,,
\end{equation}
where $\beta=8 \pi G$, the stress--energy tensor for phantom field $T_{\mu \nu}^{\text{ph}}$ and YMH $T_{\mu \nu}^{\text{YMH}}$ are respectively,  given by
\begin{align}
 T_{\mu \nu}^{\text{ph}} &= \frac{1}{2} g_{\mu \nu} \partial_\alpha \psi \partial^\alpha \psi -  \partial_\mu \psi \partial_\nu \psi \,, \\
 T_{\mu \nu}^{\text{YMH}} &= \text{Tr} \left( \frac{1}{2} D_\mu \Phi D_\nu \Phi - \frac{1}{4} g_{\mu \mu} D_\alpha \Phi D^\nu \Phi      \right) + 2 \text{Tr} \left(  g^{\alpha \beta} F_{\mu \alpha} F_{\nu \beta} - \frac{1}{4} g_{\mu \nu} F_{\alpha \beta} F^{\alpha \beta}   \right) \,.
\end{align}
The equations of motion for the matter fields are
\begin{equation} \label{matter_eqn}
  \frac{1}{\sqrt{-g}}  \partial_\mu \left(  \sqrt{-g} \partial^\mu \psi  \right)  = 0 \,,  \quad D_\mu  F^{\mu \nu}=\frac{i e}{4} \left[ \Phi, D^{\nu} \Phi   \right]  \,, \quad D_\mu   D^\mu \Phi = 0  \,.
\end{equation}
%\begin{align} 
 %  \frac{1}{\sqrt{-g}}  \partial_\mu \left(  \sqrt{-g} \partial^\mu \psi  \right)  &= 0 \,,  \label{KGeqn}  \\
%\nabla_\mu F^{\mu \nu} + i e \left[ A_\mu, F^{\mu \nu} \right] -\frac{i e}{4} \left[ \Phi, D^{\nu} \Phi   \right] &= 0 \label{YMeqn}  \,,  \\
%\nabla_\mu D^\mu \Phi + i e \left[ A_\mu, D^\mu \Phi \right] - \lambda (\Phi^2-\eta^2) \Phi &= 0 \,, \label{Higgseqn}
%\end{align}
%where $D_\mu=\nabla_\mu+ i e [A_\mu,\cdot]$.

We employ the following line element to construct a globally regular wormhole spacetime,
\begin{equation}  \label{line_element}
 ds^2 = - F_0(\eta) dt^2 + F_1(\eta) \left[  d \eta^2  +  h(\eta) \left( d \theta^2+\sin^2 \theta d\varphi^2 \right) \right]  \,,
\end{equation}
where $h(\eta)=\eta^2+\eta_0^2$ with $\eta_0$ as the throat parameter. The wormhole spacetime possesses two asymptotically flat regions in the limit $\eta \rightarrow \pm \infty$. 

Likewise, we employ the spherically symmetric Ansatz in purely magnetic gauge field $(A_t=0)$ for the gauge and Higgs field \cite{Brihaye:1998cm}, 
\begin{equation} \label{gaugefield}
 A_{\mu} dx^\mu = \frac{1-K(\eta)}{2 e} \left(  \tau_{\varphi}  d \theta -   \tau_\theta  \sin \theta d\varphi  \right) \,, \quad \Phi =  H(\eta) \tau_\eta  \,.
\end{equation}

\subsection{Ordinary Differential Equations (ODEs)}

We set the constant $e=1$. By substituting Eqs. \eqref{line_element} and \eqref{gaugefield} into the Einstein equation Eq. \eqref{einstein_eqn} and equations of motion for the gauge fields Eq. \eqref{matter_eqn}, we obtain a set of second order and nonlinear ODEs for the metric functions,
\begin{align}
F''_1 + \frac{2 \eta}{h} F'_1 -  \frac{3 F'^2_1}{4 F_1} +  \frac{\eta_0^2 F_1 }{h^2}   &=  \beta \frac{F_1}{2} \psi'^2 - \beta \frac{(K^2-1)^2 + 2 h K'^2 + h^2 F_1 H'^2 + 2 h F_1 H^2 K^2}{ 2 h^2 }     \,, \label{ode1} \\
 \left(  \frac{F'_1}{2 F_1} + \frac{\eta}{h} \right) \frac{F'_0}{F_0}   + \frac{F'^2_1}{4 F^2_1}  +  \frac{\eta}{h F_1} F'_1  -   \frac{\eta_0^2}{h^2 }  
&= - \frac{ \beta}{2} \psi'^2  + \beta  \frac{ - (K^2-1)^2 + 2 h K'^2 + h^2 F_1 H'^2 - 2 h F_1 H^2 K^2}{2 h^2 F_1}   \,,  \label{ode2}  \\
F''_0 + \left( - \frac{F'_0}{2 F_0} + \frac{\eta}{h} \right) F'_0 +  \frac{F_0}{F_1} F''_1 +  \left( - \frac{F'_1}{F_1} + \frac{\eta}{h} \right) \frac{F_0 F'_1}{F_1}  & + \frac{2 F_0 \eta_0^2}{h^2}    \nonumber \\
&=  \beta F_0 \psi'^2  -  F_0 \left[ \frac{-(K^2-1)^2 + h^2 F_1 H'^2}{ h^2 F_1}  \right]    \,. \label{ode3}  \\
 K'' + \frac{1}{2} \left(  \frac{F'_0}{F_0} - \frac{F'_1}{F_1}    \right) K' - \frac{K (K^2-1+ h F_1 H^2)}{h} &= 0 \,,  \label{odeYM}  \\
 H'' + \frac{1}{2} \left(  \frac{F'_0}{F_0} + \frac{F'_1}{F_1} + \frac{4 \eta}{h}   \right) H' - \frac{2 K^2}{h}  &= 0 \,. \label{odeHiggs} 
\end{align}
where the prime denotes the derivative of the functions w.r.t. the radial coordinate $\eta$.

We obtain a first order integral from Eq. \eqref{matter_eqn} for the phantom field, 
\begin{equation} \label{odepsi}
 \psi' =  \frac{D}{h \sqrt{F_0 F_1}} \,,
\end{equation}
where $D$ is the scalar charge of the phantom field. Then we replace the term $\psi'^2$ by substituting $\psi'=D/(h\sqrt{F_0 F_1})$ into the Eqs.~\eqref{ode1}-\eqref{ode3}.

We solve Eqs. \eqref{ode1}, \eqref{ode3}, \eqref{odeYM} and \eqref{odeHiggs} numerically with Eq. \eqref{ode2} is expressed as 
%\begin{align}
% D^2 &=  \frac{2 h^2 F_0 F_1}{\beta} \left[   - \left(  \frac{F'_1}{2 F_1} + \frac{\eta}{h} \right) \frac{F'_0}{F_0}   - \frac{F'^2_1}{4 F^2_1} -  \frac{\eta}{h F_1} F'_1 +  \frac{\eta_0^2}{h^2 }  \nonumber  \right. \\
%& \qquad  \left. + \beta  \frac{ - (K^2-1)^2 + 2 h K'^2 + h^2 F_1 H'^2 - 2 h F_1 H^2 K^2}{2 h^2 F_1}    \right] \,. \label{Constr}
%\end{align}
\begin{equation}
  D^2 =  \frac{2 h^2 F_0 F_1}{\beta} \left[   - \left(  \frac{F'_1}{2 F_1} + \frac{\eta}{h} \right) \frac{F'_0}{F_0}   - \frac{F'^2_1}{4 F^2_1} -  \frac{\eta}{h F_1} F'_1 +  \frac{\eta_0^2}{h^2 } + \beta  \frac{ - (K^2-1)^2 + 2 h K'^2 + h^2 F_1 H'^2 - 2 h F_1 H^2 K^2}{2 h^2 F_1}    \right] \,. \label{Constr}
\end{equation}
to monitor quality of the numerical solutions with the condition $D^2=const$.

\subsection{Geometrical Properties}

We introduce $R(\eta)^2$ as the shape function for wormholes to study the geometry of wormhole,
\begin{equation}
 R(\eta)^2 = F_1 h \,.
\end{equation} 
Note that $R(\eta)$ is the circumferential radius of the wormhole and should not contain zero for a globally regular wormhole solution. When $R$ contains a local minimum, which is known as the throat of wormhole, then the wormhole possesses a minimal surface area at the throat. However, if $R$ contains a local maximum, then it is an equator of the wormhole, which corresponds to the maximal surface area of wormhole. 

For simplicity, we consider the metric functions symmetric w.r.t. the coordinate $\eta=0$, so we assume the circumferential radius of wormhole at $\eta=0$ could be either a throat or an equator, which implies $R$ should have an extremum at $\eta=0$ by requiring 
\begin{equation}
  R'(0) = 0 \quad \Rightarrow \quad \frac{\left( h F'_1 + 2 \eta F_1 \right)}{2 R}   \Bigg|_{\eta=0} = 0\,,
\end{equation}
where we have to set $F'_1(0)=0$. In particular, if the wormhole only contains a single throat, then the throat must be located at $\eta=0$ with the minimal surface area $A_{\text{th}}= 4 \pi R(0)^2=4 \pi F_1(0) \eta_0^2$. 

Furthermore, the second order derivative of $R$ at $\eta=0$ is given by 
\begin{equation}
R''(0) =  \frac{2 F_1+h F''_1}{2 R} \Bigg|_{\eta=0} = \frac{F_1}{R}  \,,
\end{equation}
where we have used $F''_1(0)$ from the ODEs. The wormholes possess either a throat or an equator at $\eta=0$, which that can be determined, respectively from the conditions $R''(0) > 0$ and $R''(0) <0$. Since $R''(0)>0$, then $R(0)$ always stays as throat. Note that when $R'(\eta_\text{crit})=R''(\eta_\text{crit})=0$, the circumferential radius forms a turning point at some value of the radial coordinate $\eta_\text{crit}$, the geometry of wormhole is in a transition state which the double throat and the equator can simultaneously exist, this also implies that there is a transition can occur from the single throat configuration to the double throat configuration \cite{Dzhunushaliev:2014bya,Hoffmann:2017jfs}.

In addition, we can visualize the wormhole throat in the equatorial plane $(\theta=\pi/2)$ by embedding the equatorial plane to the cylindrical coordinates $(\rho,\varphi,z)$ in Euclidean space, 
\begin{align}
ds^2 &= F_1  d \eta^2 + h F_1  d \varphi^2 \,, \\
&= d \rho^2 + dz^2 + \rho^2 d \varphi^2   \,.
%&=   \left[ 1+ \left(  \frac{dz}{d \rho}   \right)^2    \right]  d \rho^2 + \rho^2 d \varphi^2  \,. 
\end{align}
Hence, we obtain the expression for $z$ from the comparison,
\begin{equation} \label{formula_embedding}
  z =  \pm \int  \sqrt{  F_1  -   \left( \frac{d \rho}{d \eta} \right)^2    }     d \eta \,, \quad \rho \equiv R \,. 
\end{equation}
%The sign of $z$ depends on the sign of the radial coordinate $r$. Note that we can obtain an analytical expression of $z$ only for Morris--Throne type wormhole in GR which is massless and symmetric $(A=N=1)$,
%\begin{equation}
%5z &= \int_{0}^{l} \sqrt{  1  -   \left( \frac{d R}{d l'} \right)^2    }     d l'   \nonumber   \\
%z =  \int_{0}^{\eta} \sqrt{  1  -  \frac{{r'}^2}{{r'}^2+{r}_0^{2}}     }   \,  d r' = \text{arcsinh} \left( \frac{r}{r_0} \right) \,.
%\end{equation}

We can also evaluate the surface gravity $\kappa$ at the throat, which is defined as
\begin{align} 
 \kappa^2 &= -\frac{1}{2} \left(   \nabla_\mu \zeta_\nu \right) \left(   \nabla^\mu \zeta^\nu   \right)    \,,  \\
\Rightarrow \kappa &= \frac{F'_0}{2 \sqrt{F_0 F_1} } \,, \label{surgra}
\end{align}
where $\zeta^\mu=(1,0,0,0)$ is the timelike Killing vector. Eq. \eqref{surgra} shows that $\kappa$ vanishes for symmetric wormholes with a single throat but remains finite for wormholes with double throat configuration.

\subsection{Global Charges}

The wormhole solutions possess mass $M$ as the global charge associated with the asymptotic Killing vector $\partial_t$. The mass of wormhole can be read off directly from the asymptotic expansion of the metric at $\eta \rightarrow \infty$,
\begin{equation}
 F_0 \rightarrow 1- \frac{2 G M}{\eta} \,.
\end{equation}
%where the mass parameter $\mu$ is given by
%\begin{equation}
% \mu = \frac{\beta M}{8 \pi \eta_0} \,.
%\end{equation}

Recall that the charge of the phantom field is given by $D^2$. Then the magnetic charge for the non--abelian gauge fields is given by \cite{Corichi:2000dm,Ashtekar:2000nx}
\begin{equation}
 \mathcal{P}^{\text{YMH}} = \frac{1}{4 \pi} \oint \sqrt{ \sum_i \left(  F^i_{\theta \varphi} \right)^2    } d\theta d\varphi = |P| \,,
\end{equation}
where the integral is evaluated at the spatial infinity, yielding $P=0$ for the hairy wormholes \cite{hauser2014hairy}.

\subsection{Null Energy Condition (NEC)}
Since the construction of wormhole requires the violation of energy conditions. Then, we can examine the NEC of the wormhole, which states that
\begin{equation}
  T_{\mu \nu} k^\mu k^\nu \geq 0 \,,
\end{equation}
for all (future--pointing) null vector $k_\mu$. Note that the violation of NEC also implies the violation of the weak and strong energy conditions.

Since the wormhole spacetime is spherically symmetric, there are two choices of null vector \cite{Antoniou:2019awm}, 
\begin{equation}
k_\mu = \left( g_{tt} , \sqrt{-\frac{g_{tt}}{g_{\eta \eta}}}, 0, 0 \right) \,, \quad \text{and} \quad k_\mu = \left( 1, 0, \sqrt{- \frac{g_{tt}}{g_{\theta \theta}}  } ,  0 \right)\,,
\end{equation}
which yield two expressions to measure the violation of NEC,
%\begin{equation}
% -T^t\,_t + T^\eta\,_\eta \geq 0 \,, \quad  -T^t\,_t + T^\theta\,_\theta \geq 0 \,.
%\end{equation}
%The above expressions can be evaluated explicitly, 
\begin{align}
-T^t\,_t + T^\eta\,_\eta &=  - \frac{\psi'^2}{F_1} + \frac{H'^2}{F_1} + \frac{2 K'^2}{h F_1^2} =  - \frac{D^2}{h^2 F_0 F_1^2} + \frac{H'^2}{F_1} + \frac{2 K'^2}{h F_1^2} \,,   \\
  -T^t\,_t + T^\theta\,_\theta &=   \frac{K'^2}{h F_1^2} + \frac{H^2 K^2}{h F_1} + \frac{\left( K^2-1 \right)^2}{h^2 F_1^2} \geq 0 \,.
\end{align}

\subsection{Boundary Conditions}

Since we only consider the wormhole solutions with the metric functions symmetric w.r.t. $\eta=0$, thus we only integrate the ODEs from $\eta=0$ to the infinity. We impose the eight boundary conditions at $\eta=0$ and $\eta=\infty$. First, we require the first order derivative of the metric functions vanish at the throat, 
\begin{equation}
 F'_0 (0) = F'_1 (0) = 0 \,.
\end{equation}
These imply that the metric functions possess the exremum at $\eta=0$. At the infinity, the metric functions approach Minkowski spacetime,
\begin{equation}
 F_0 (\infty) = F_1 (\infty) = 1 \,.
\end{equation}

We impose the following boundary conditions for the gauge fields, by fixing their values at $\eta=0$ and they satisfy the asymptotic flatness at the infinity,
\begin{equation}
 K(0)=1 \,, \quad H(0) = 0\,, \quad K(\infty) = 0 \,, \quad H(\infty)=1 \,.
\end{equation}

\begin{figure}[t!]
\centering
\mbox{
(a)
 \includegraphics[angle =-90,scale=0.3]{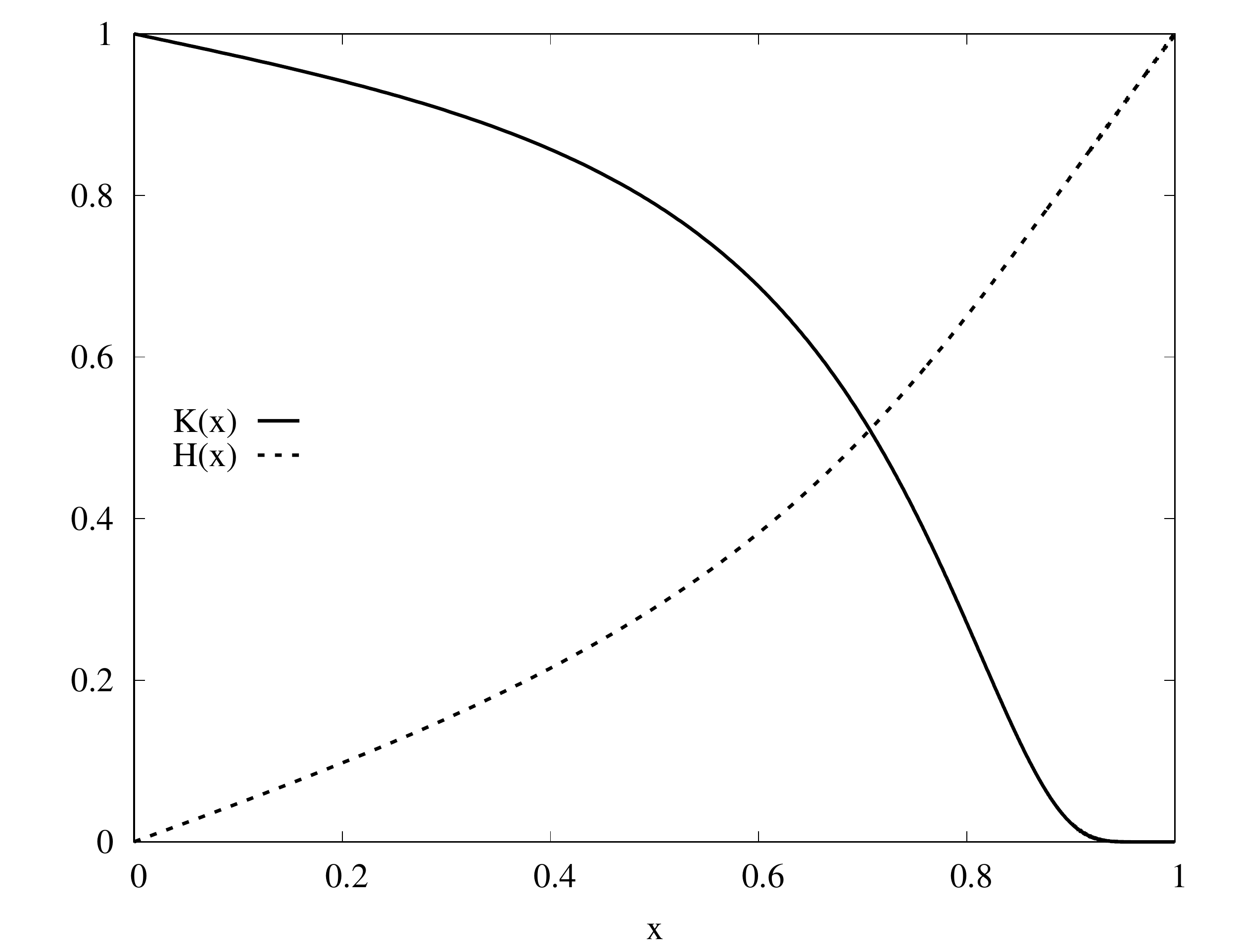}
%(b)
% \includegraphics[angle =-90,scale=0.3]{plot_HooftPolyakov_Mass}
 }
%\mbox{
%(c)
% \includegraphics[angle =-90,scale=0.3]{plot_HooftPolyakov_Mass}
%}
\caption{The gauge fields $K(x)$ and $H(x)$ in the compactified coordinate $x$ in the probe limit.}
\label{plot_ProbeLimit}
\end{figure}

We solve the set of ODEs numerically by Colsys which solves boundary value problems for systems of nonlinear coupled ODEs based on the Newton--Raphson method \cite{colsys}. We scale the parameters by the throat parameter $\eta_0$,
\begin{equation}
 \eta \rightarrow \eta_0 \eta \,, \quad  h \rightarrow \eta_0^2 h  \,, \quad  \beta \rightarrow  \eta^2_0 \beta \,, \quad H \rightarrow \frac{ H}{\eta_0} \,. 
\end{equation}
%\begin{equation}
% \eta =\eta_0  \xi \,, \quad  h= \eta_0^2 \bar{h} = \eta_0^2 ( \xi^2 +1) \,, \quad  \beta = \eta^2_0 \bar{\beta} \,, \quad H =\frac{ \bar{H}}{\eta_0} \,. 
%\end{equation}
Thus we introduce a mass parameter $\mu$, which is given by
%Thus the mass $M$ can be scaled by
%\begin{equation}
% F_0 \rightarrow  1 - \frac{2 \mu}{\xi}\,,
%\end{equation}
%where  is given by 
\begin{equation}
 \mu = \frac{\beta M}{8 \pi \eta_0} \,.
\end{equation}
We compactify the radial coordinate $\eta$ by $\eta= \eta_0 \tan \left( \pi x/2 \right)$ in the numerics. 
%We compactify the radial coordinate $\xi$ by $\xi=  \tan \left( \pi x/2 \right)$ in the numerics. %We find that the ODEs possess the following scaling symmetry,
%\begin{equation}
% \eta \rightarrow  \frac{\eta}{\gamma} \,, \quad \eta_0 \rightarrow \frac{\eta_0}{\gamma}\,, \quad \beta \rightarrow  \frac{\beta}{\gamma^2} \,, \quad H \rightarrow \gamma H\,, \quad \upsilon \rightarrow \gamma \upsilon \,. 
%\end{equation}

\section{Results and Discussions}

%\begin{figure}[t!]
%\centering
%\mbox{
%(a)
% \includegraphics[angle =-90,scale=0.3]{plot_HooftPolyakov_KH}
%(b)
% \includegraphics[angle =-90,scale=0.3]{plot_HooftPolyakov_Mass}
 %}
%\mbox{
%(c)
% \includegraphics[angle =-90,scale=0.3]{plot_HooftPolyakov_Mass}
%}
%\caption{The solutions of Hooft--Polyakov monopole in the compactified coordinate, $x=R/(1+R)$: (a) $K(x)$ and $H(x)$ (b)  The mass of Hooft--Polyakov Monopole as a function of $\lambda$}
%\label{plot_HP}
%\end{figure}

\subsection{Probe Limit}

\begin{figure}[t!]
\centering
\mbox{
(a)
 \includegraphics[angle =-90,scale=0.3]{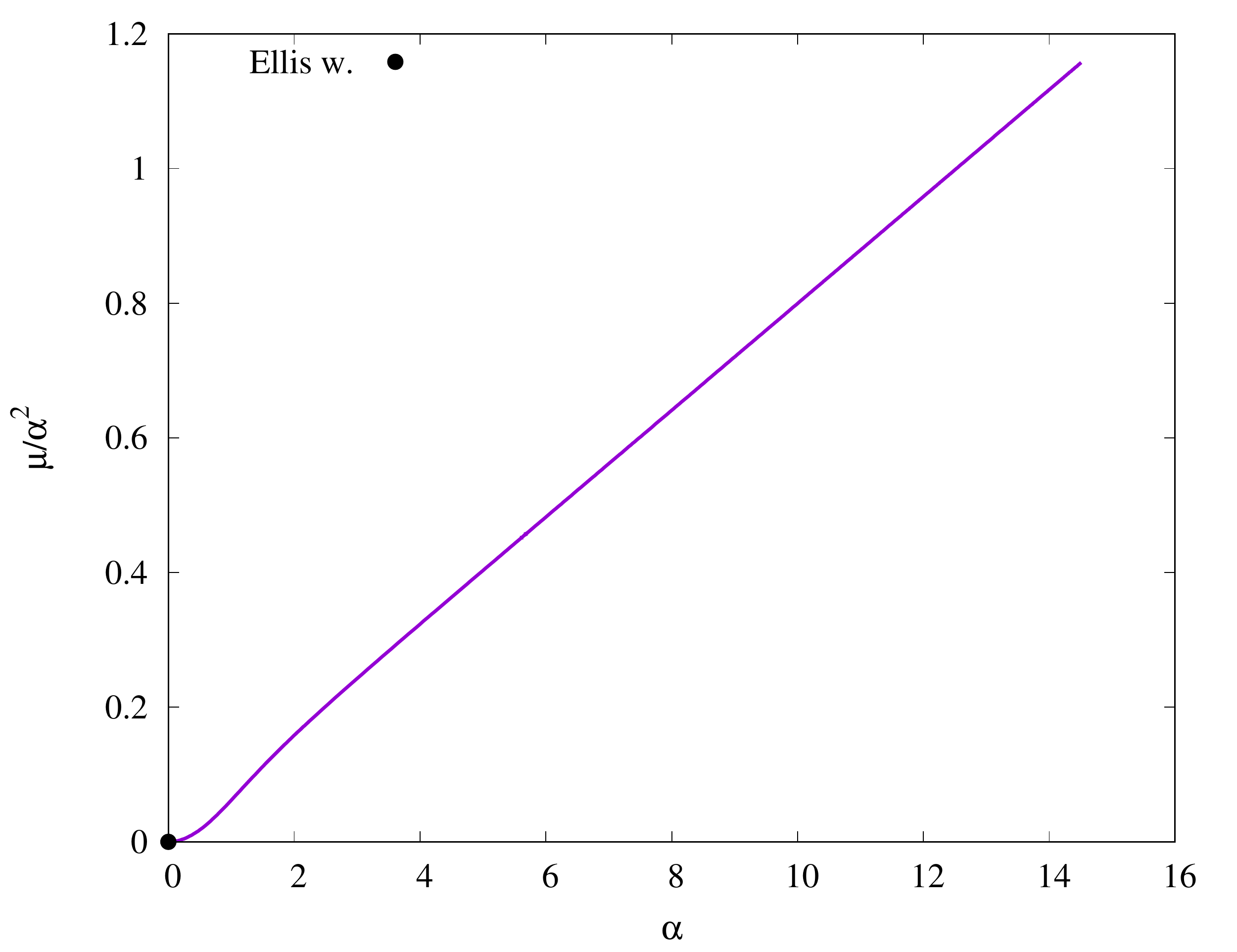}
(b)
 \includegraphics[angle =-90,scale=0.3]{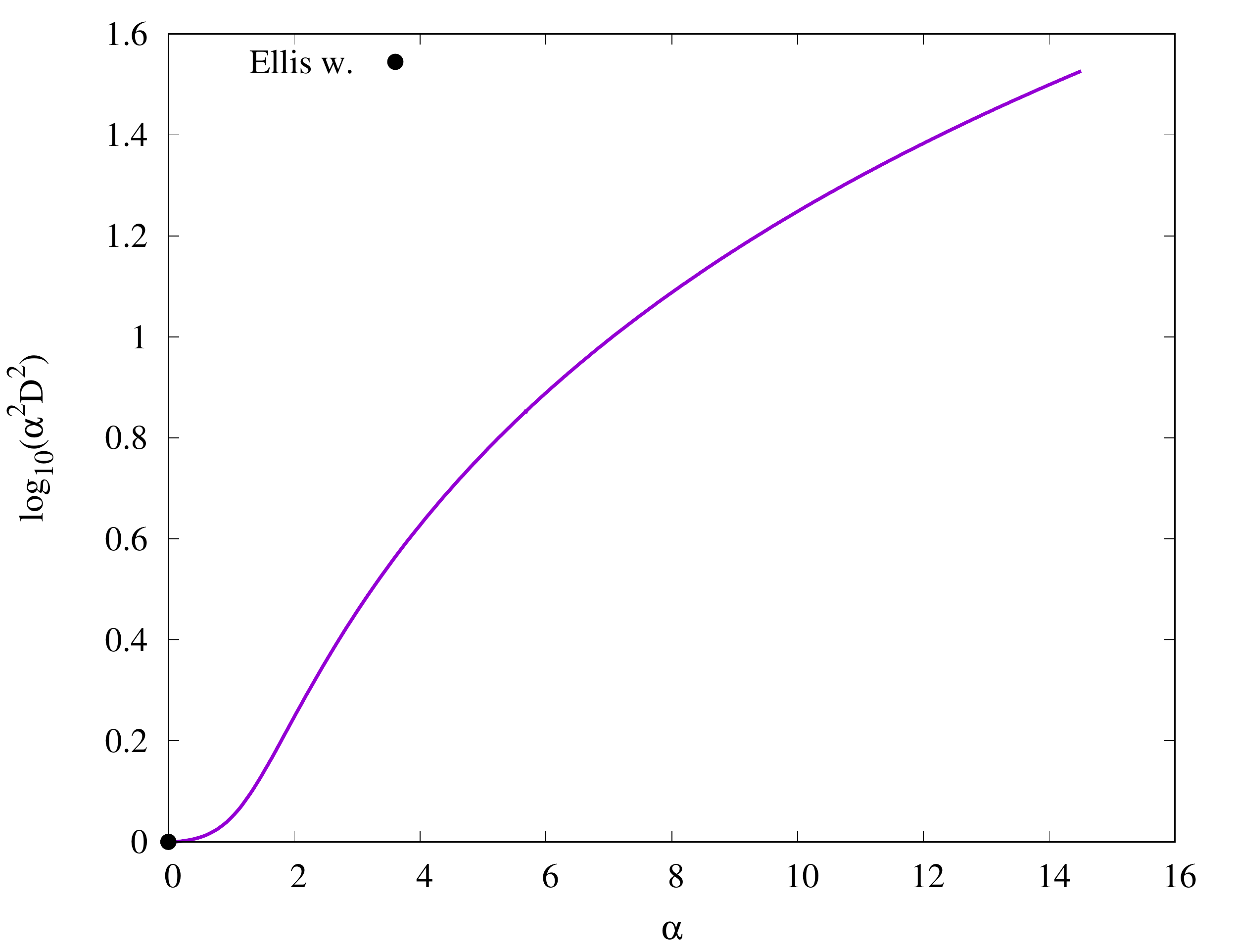}
 }
\mbox{
(c)
 \includegraphics[angle =-90,scale=0.3]{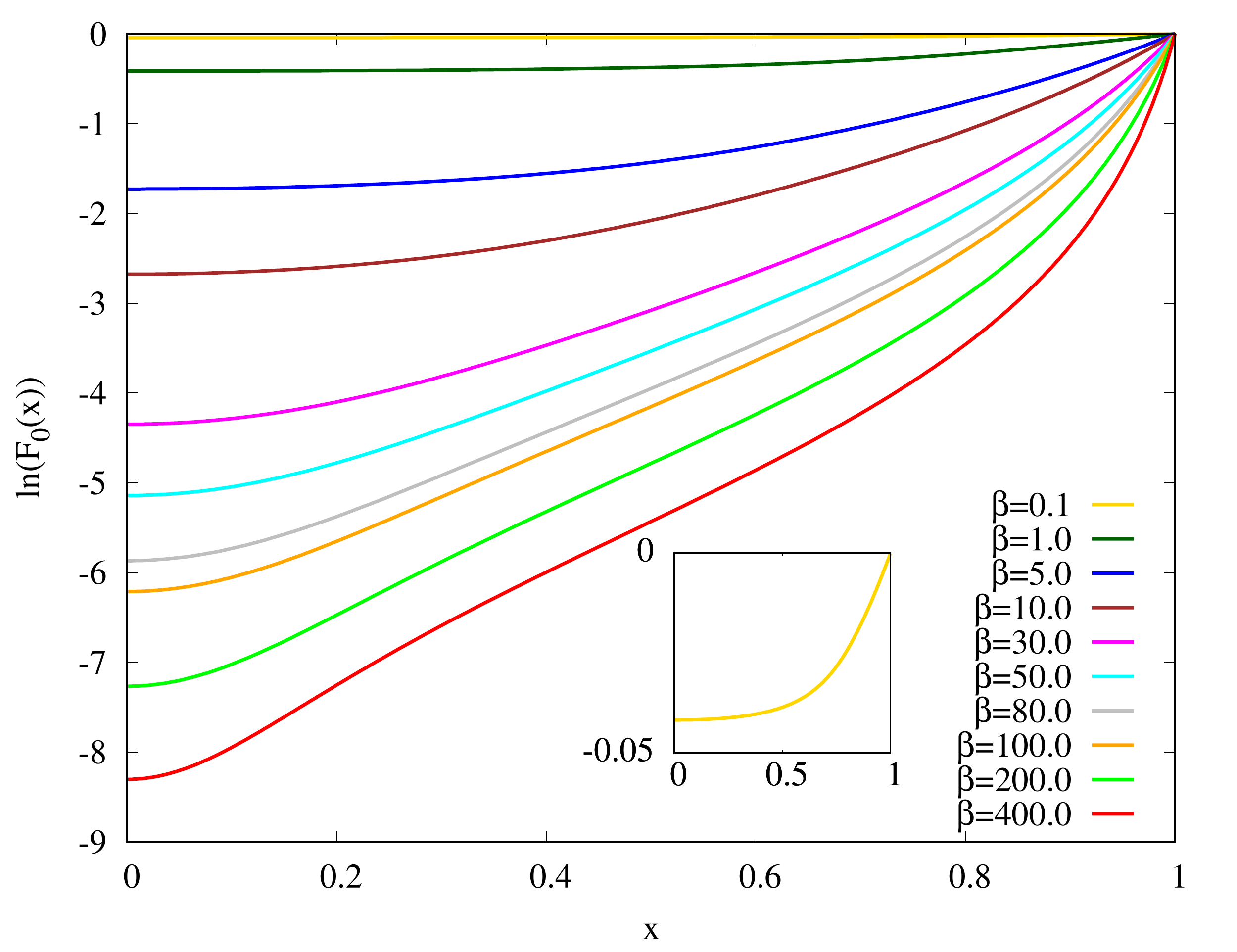}
(d)
\includegraphics[angle =-90,scale=0.3]{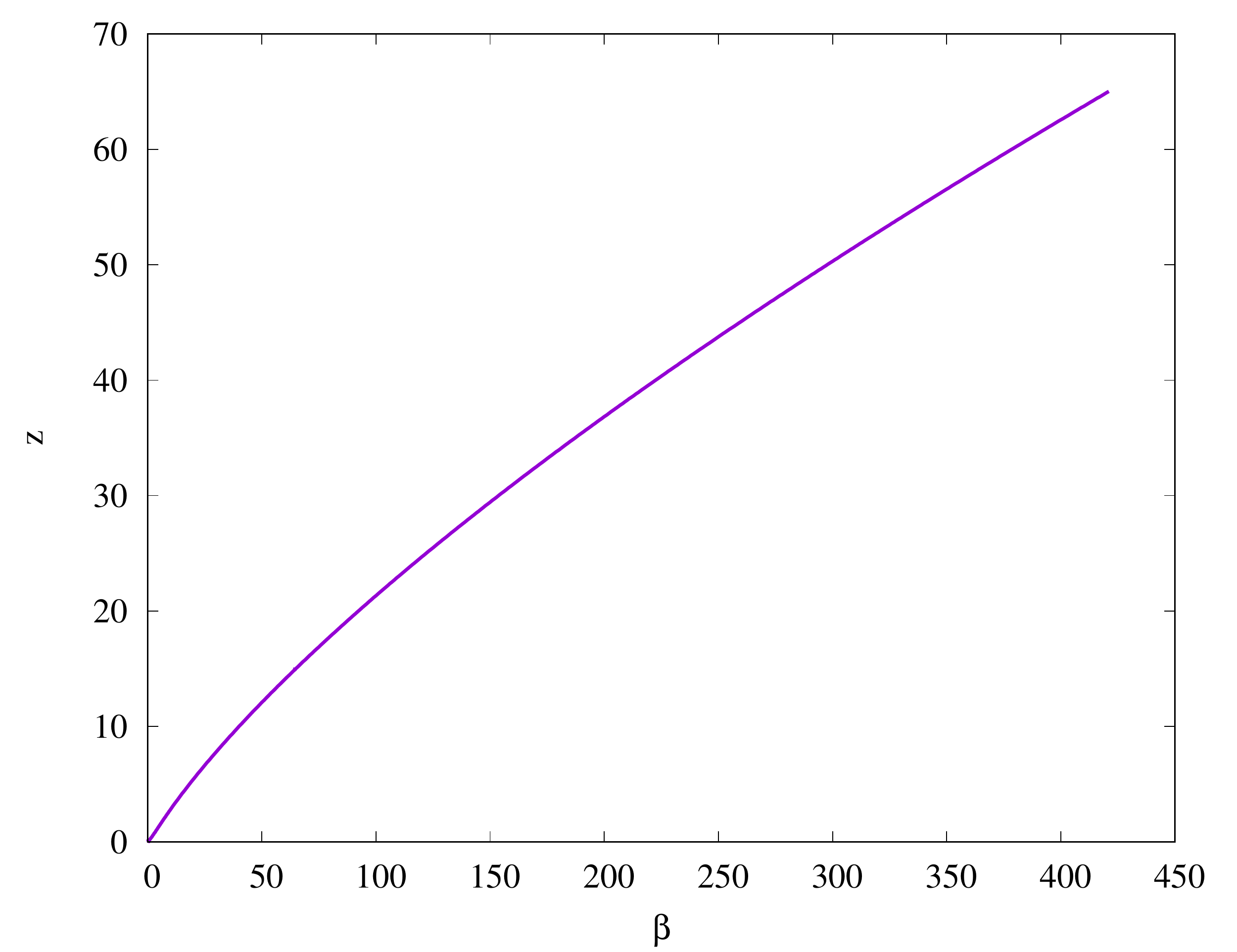}
}
\mbox{
(e)
\includegraphics[angle =-90,scale=0.3]{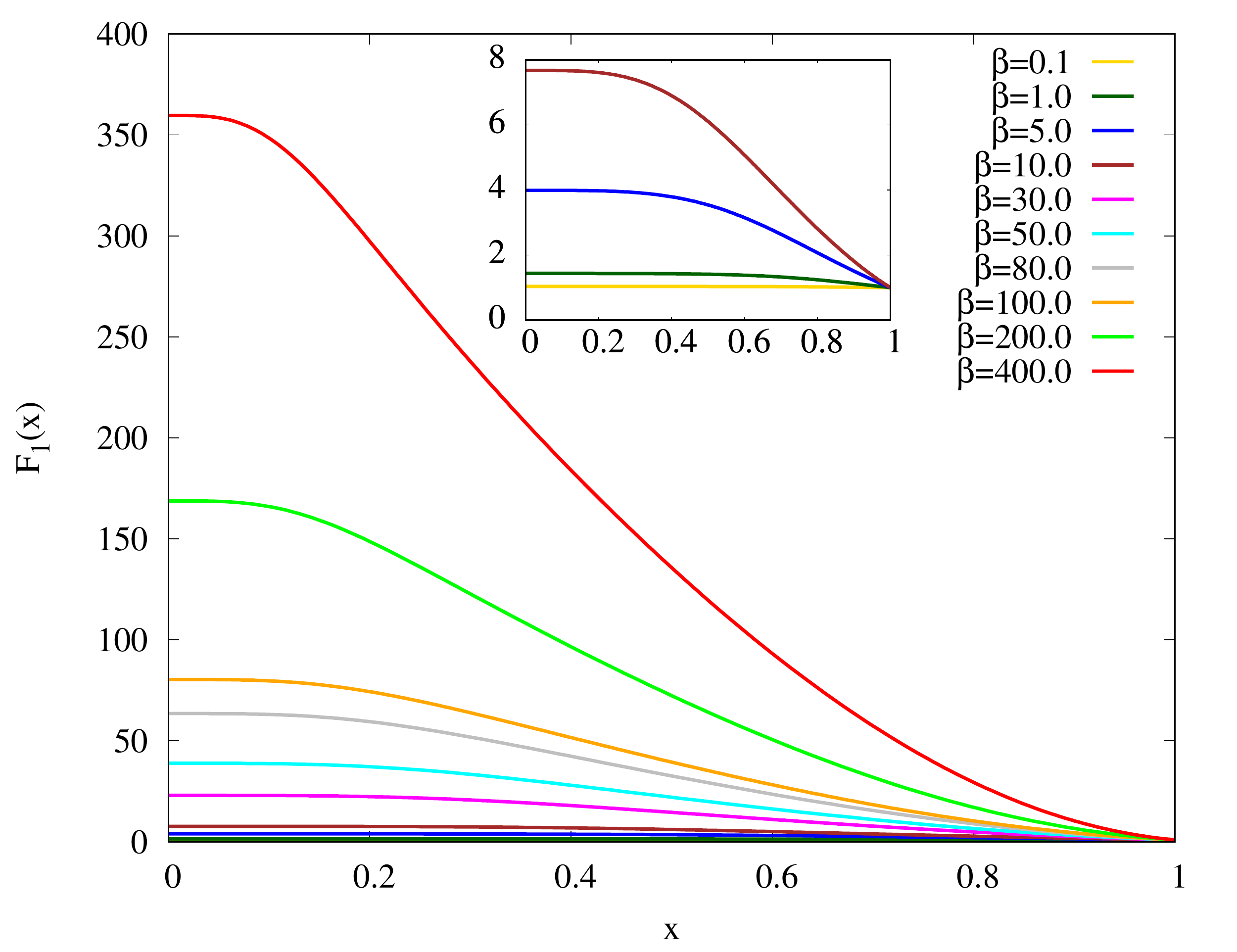}
(f)
\includegraphics[angle =-90,scale=0.3]{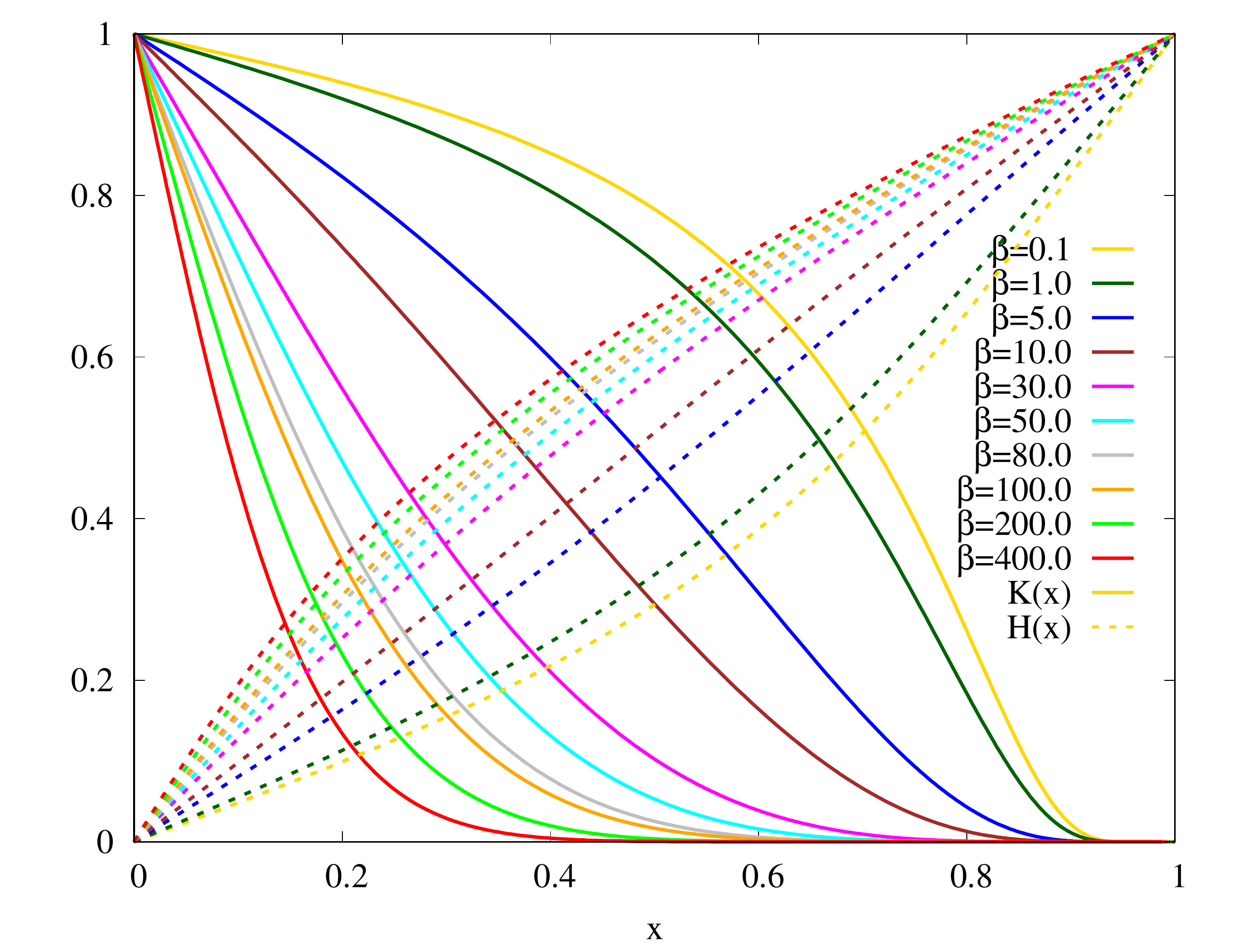}
}
\caption{ (a) The scaled mass $\mu/\alpha^2$ versus the scaled gravitational coupling constant $\alpha$; (b) The logarithmic of scaled scalar charge $\log_{10}(\alpha^2 D^2)$ versus the scaled gravitational coupling constant $\alpha$; (c) The metric function  $\ln(F_0(x))$ in the compactified coordinate $x$ for several values of $\beta$; (d) The gravitational redshift $z$ versus the gravitational coupling constant $\beta$; (e) The metric function $F_1(x)$ in the compactified coordinate $x$ for several values of $\beta$; (f) The gauge fields $K(x)$ and $H(x)$ for several values of $\beta$ in the compactified coordinate $x$; The dot denotes the value for massless Ellis wormhole.}
\label{plot_sol_l0}
\end{figure}

We start our investigation from the probe limit of the wormhole in the BPS limit. When the gravity is switched off $(\beta=0)$, we see that the YMH field doesn't contribute to the Einstein equation from Eqs.~\eqref{ode1}--\eqref{ode3}. Therefore, the metric Eq. \eqref{line_element} is the massless Ellis wormhole $ (F_0 (\eta) = F_1 (\eta)= 1)$ which is symmetric. The phantom field is given by
\begin{equation}
 \psi = \frac{D}{\eta_0} \left[  \arctan \left( \frac{\eta}{\eta_0} \right)  - \frac{\pi}{2} \right] \,.
\end{equation}

The pure YMH equations in the background of Ellis wormhole are then simplifed to
\begin{align}
 K'' &= \frac{K (K^2-1+ h H^2)}{h}  \,, \\
 H'' &= - \frac{2 \eta }{h }  H' + \frac{2 K^2}{h} H \,. 
\end{align}
%\begin{align}
% K'' &= \frac{K (K^2-1+ \bar{h} \bar{H}^2)}{\bar{h}}  \,, \\
% \bar{H}'' &= - \frac{2 \xi }{\bar{h} }  \bar{H}' + \frac{2 K^2}{\bar{h}} \bar{H} \,. 
%\end{align}
The above ODEs are solved numerically which are shown in Fig. \ref{plot_ProbeLimit}. Note that the 'Hooft--Polyakov monopole is the solutions of the YMH theory. The corresponding theory possesses an exact solution solution in the BPS limit \cite{prasad1975exact}, which is given by 
\begin{equation}
 K(R) = \frac{R}{\sinh(R)} \,, \quad H(R) = \text{coth}(R)-\frac{1}{R} \,.
\end{equation}
The exact solution is stable and its mass is unity. %\textbf{KG Maybe you can help me to veify this and put a reference for this}

\begin{figure}[t!]
\centering
\mbox{
(a)
 \includegraphics[angle =-90,scale=0.3]{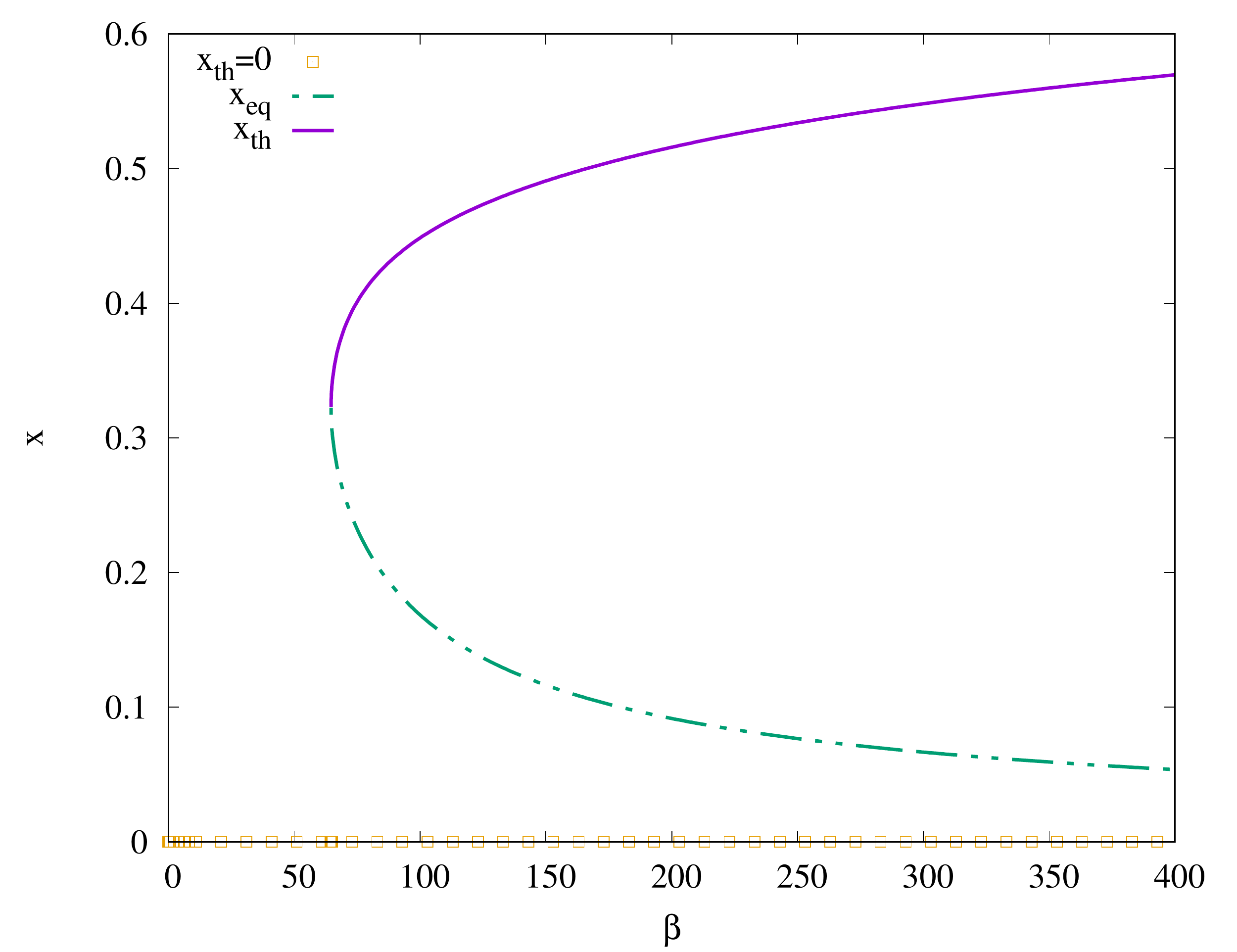}
(b)
 \includegraphics[angle =-90,scale=0.3]{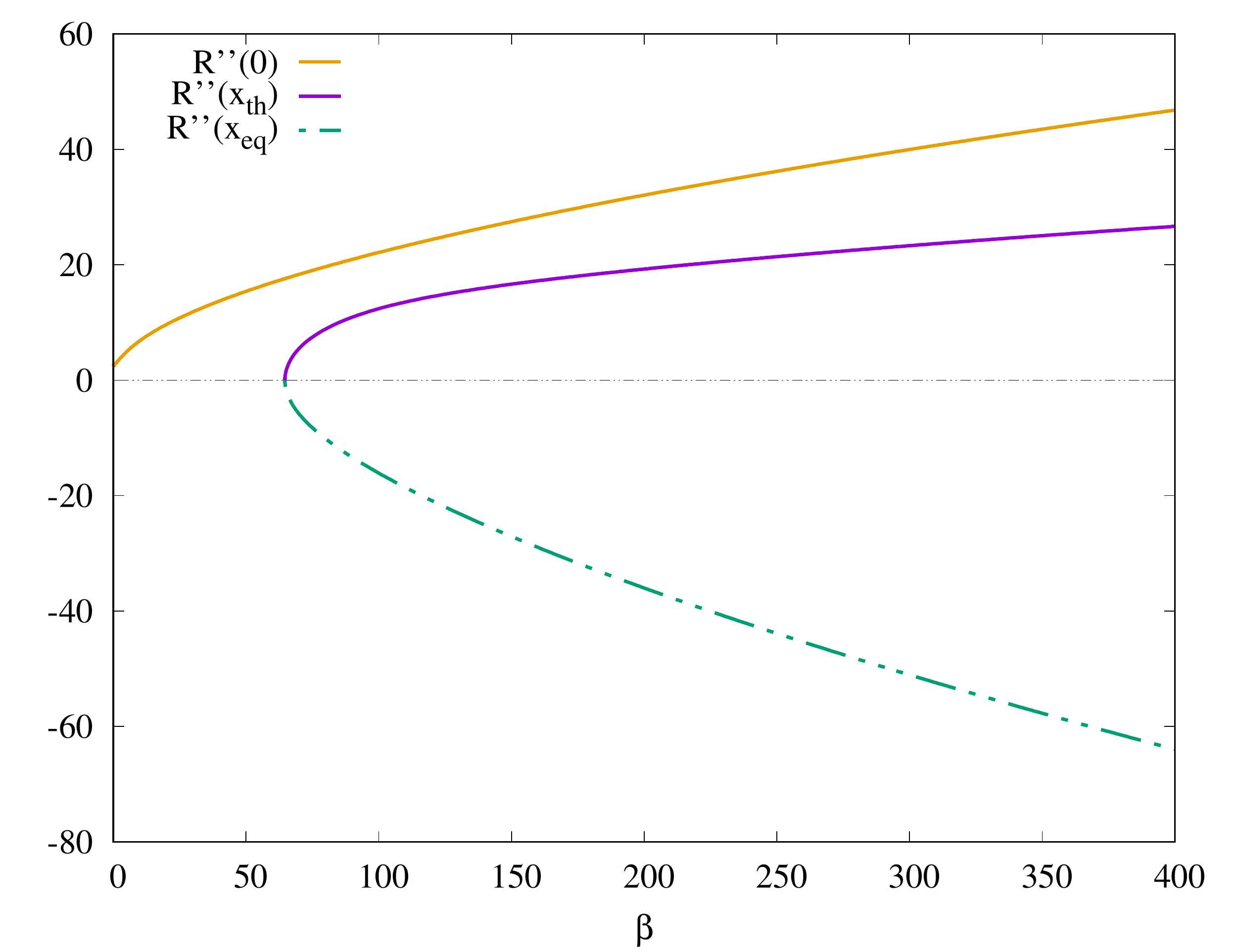}
 }
\mbox{
(c)
 \includegraphics[angle =-90,scale=0.3]{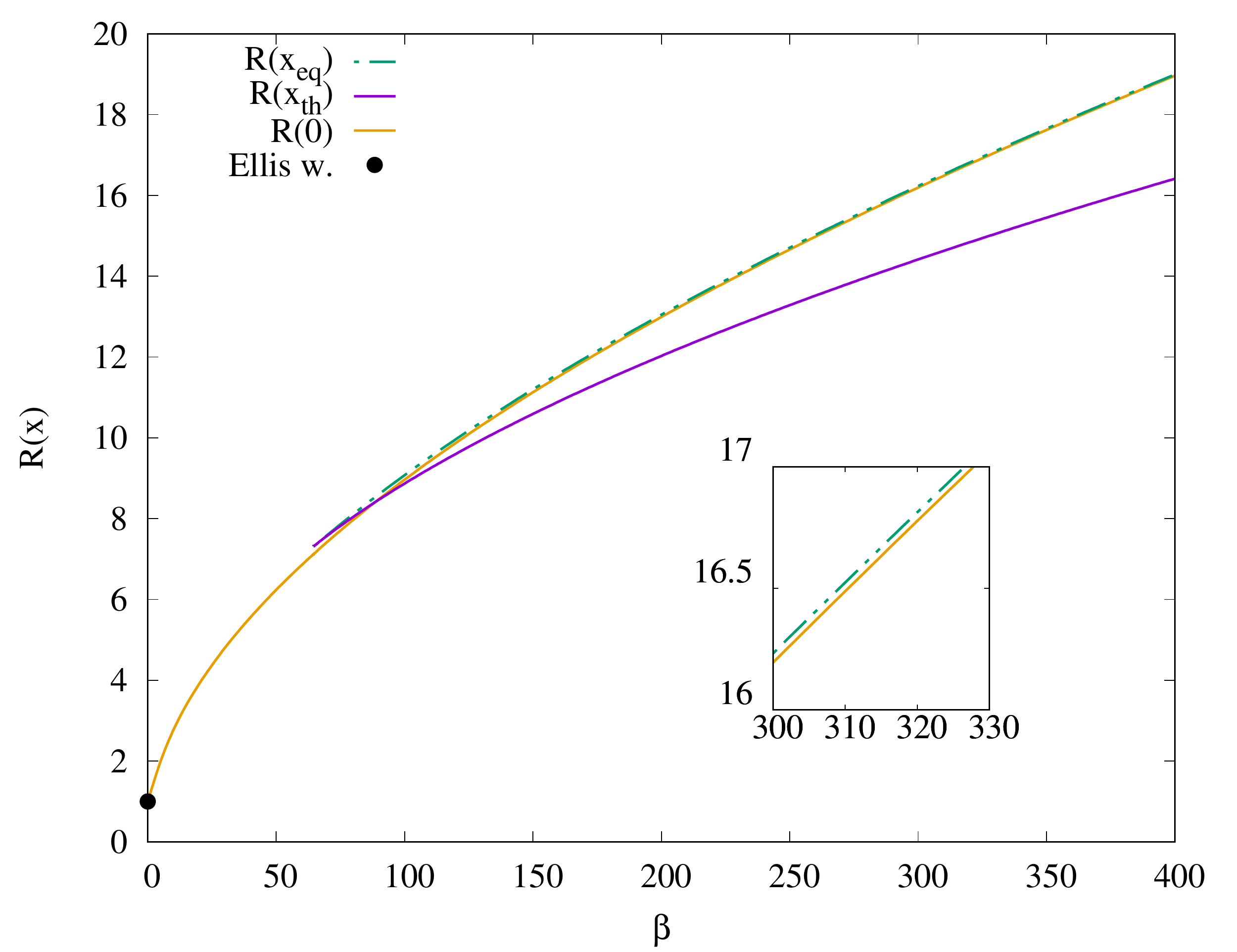}
(d)
\includegraphics[angle =-90,scale=0.3]{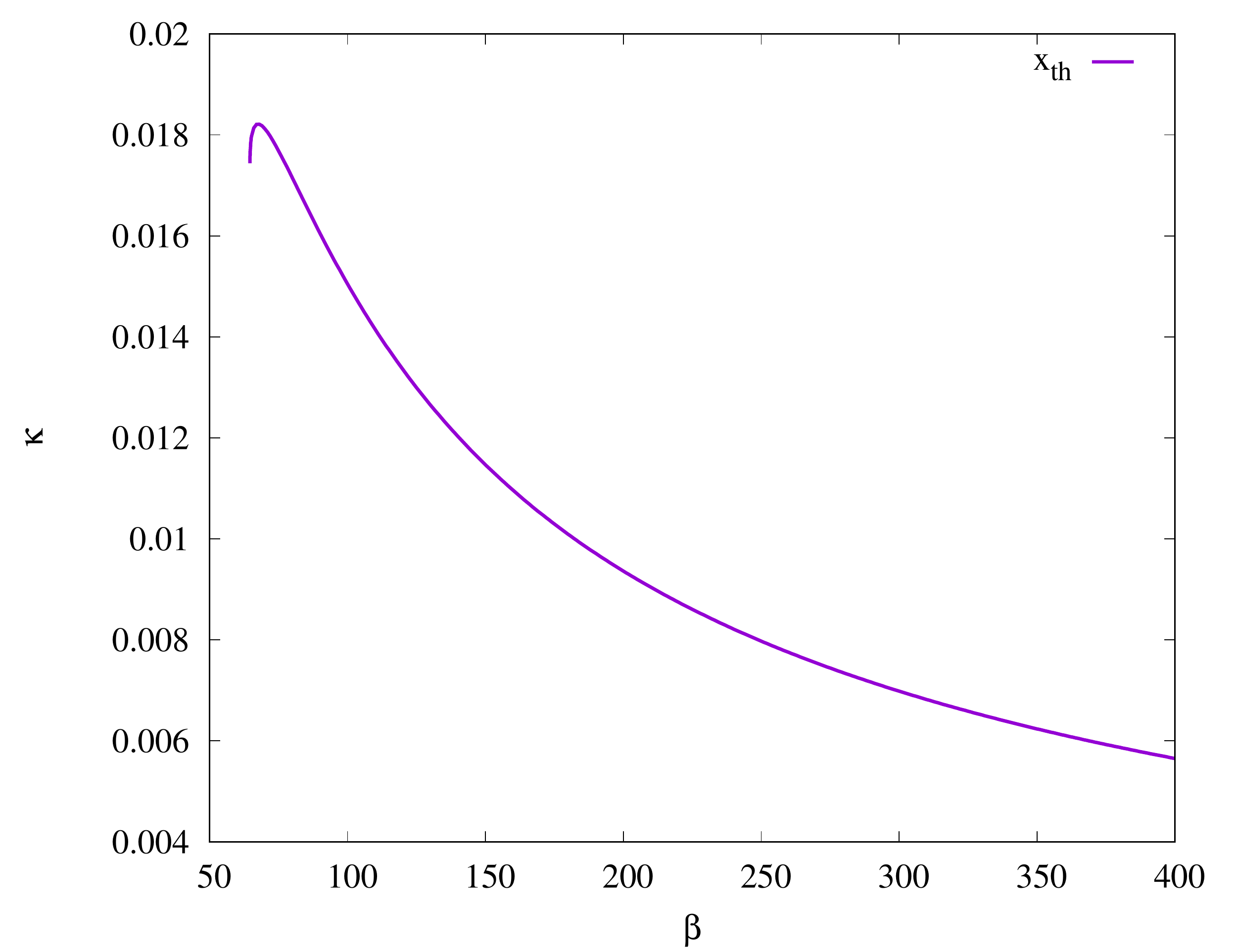}
}
\mbox{
(e)
\includegraphics[angle =-90,scale=0.3]{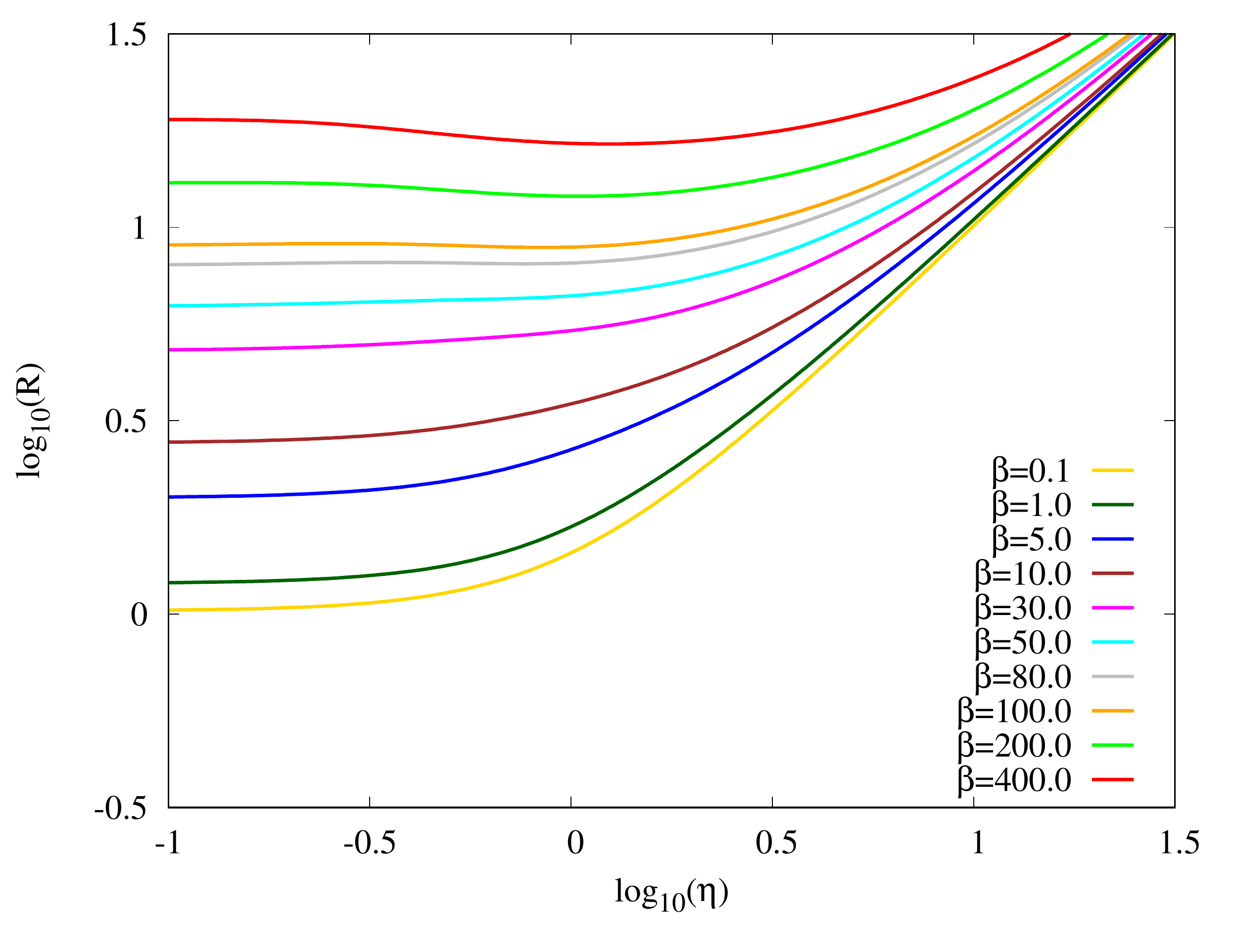}
(f)
\includegraphics[angle =-90,scale=0.3]{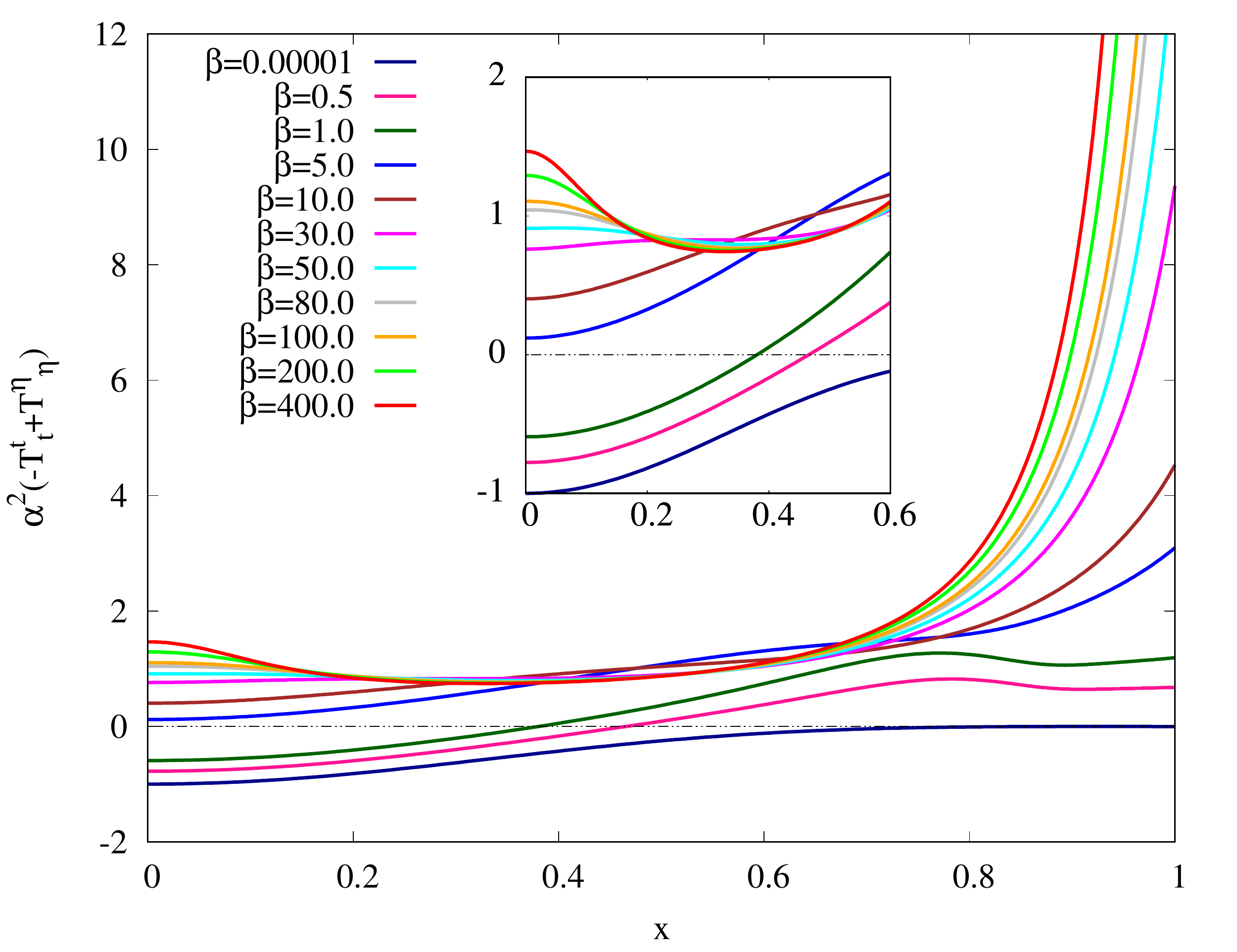}
}
\caption{The properties of wormhole solutions: (a) The location of throats and the equator versus the gravitational coupling constant $\beta$; (b) The second order derivative of $R$ in the compactified coordinate $x$ for the throats and the equator versus the gravitational coupling constant $\beta$; (c) The circumferential radius of throats $R(x_\text{th})$ and equator $R(x_\text{eq})$ in the compactified coordinate $x$ versus the gravitational coupling constant $\beta$; (d) The surface gravity $\kappa$ at the throat $x_\text{th}$  versus the gravitational coupling constant $\beta$. (e) The circumferential radius $R$ of wormhole solutions for several values of $\beta$; (f) The violation of scaled null energy condition (NEC) for the wormhole solutions with several values of $\beta$ in the compactified coordinate $x$; The constant $\alpha$ is defined as $\beta=2 \alpha^2$; In (a)--(c), the yellow curve denotes the throat at $x=0$. The green curve with two-dashed line and purple curve denote the equator $x_\text{eq}$, and another throat at $x_\text{th}$ in the compactified coordinate $x$, respectively.}
\label{plot_prop}
\end{figure}

\subsection{With Backreaction}

\begin{figure}[t!]
\centering
\mbox{
(a)
 \includegraphics[angle =0,scale=0.3]{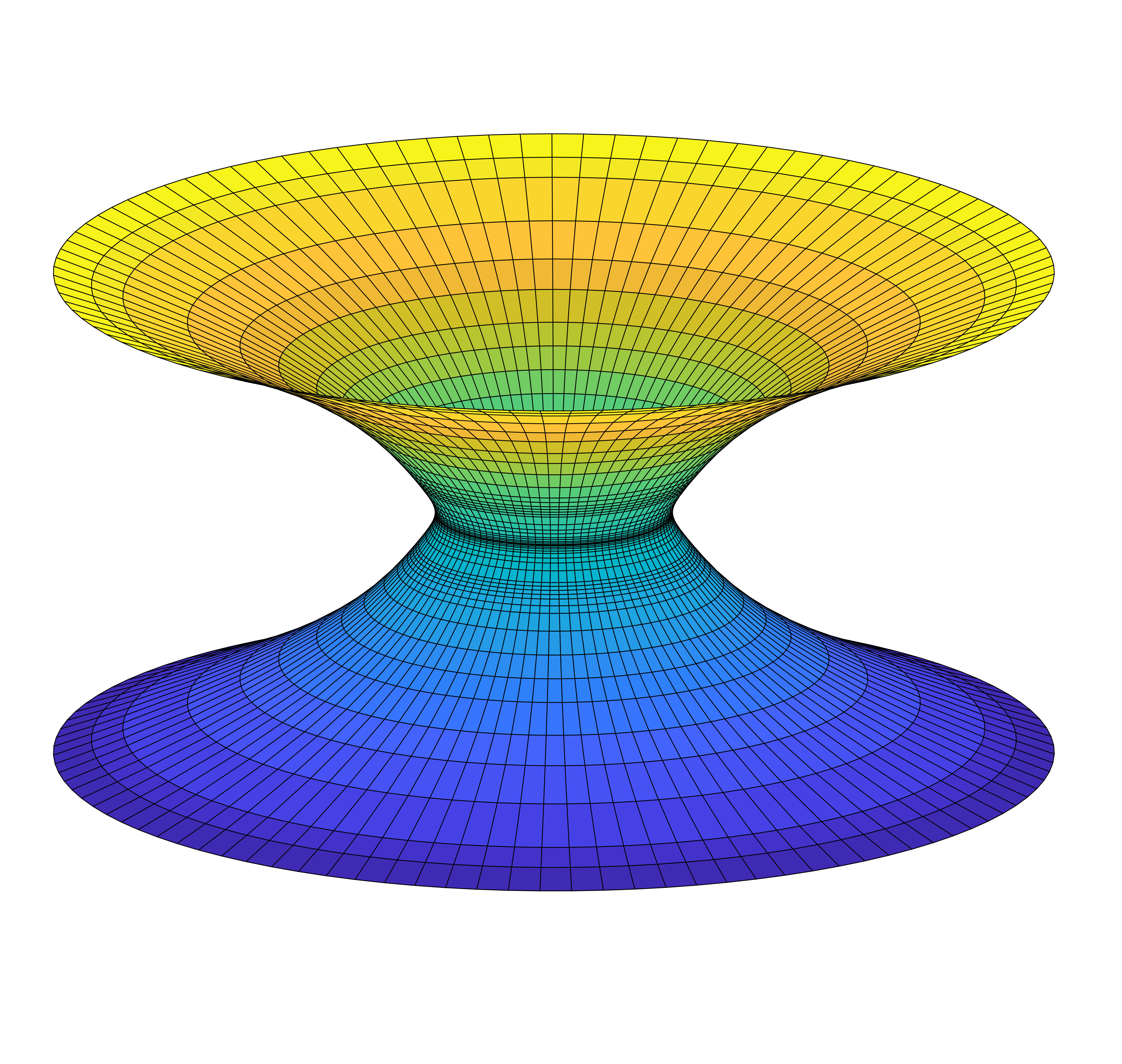}
(b)
 \includegraphics[angle =0,scale=0.3]{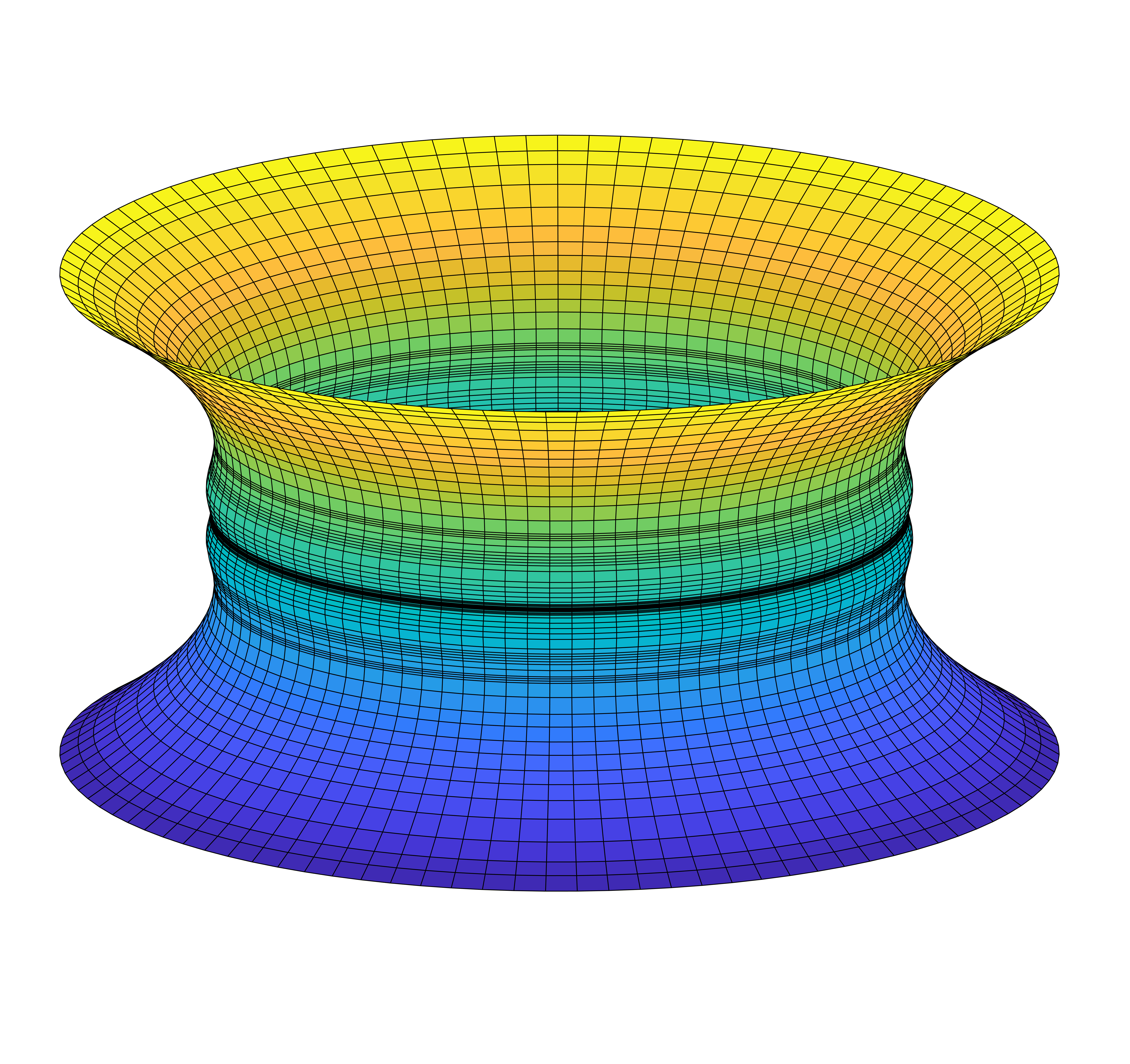}
 }
\caption{The isometric embedding of the wormholes in the Euclidean space for a) $\beta=10$ (single throat) and b) $\beta=100$ (double throat).}
\label{plot_embed3d}
\end{figure}

We exhibit our numerical results by varying the gravitational coupling constant $\beta$ in the range $[0,400]$. The wormholes could take any real positive values of $\beta$, this is contrast to YM wormhole where the limiting configuration is the extremal Reissner--Nordstrom black hole for higher nodes \cite{hauser2014hairy}. Recall that Ellis wormhole is massless, its circumferential radius and scalar charge are unity. Since the gauge fields don't present, then the Ellis wormhole has the analogue of Schwarzschild solution for the black holes \cite{hauser2014hairy}. When we increase $\beta$ from zero, the solutions of hairy wormholes emerge from the Ellis wormhole, thus the properties of these hairy wormholes differ from Ellis wormholes when $\beta$ increases. Fig. \ref{plot_sol_l0}(a) shows that the hairy wormholes gain the mass when $\beta$ increases from zero, the mass increases monotonically as $\beta$ increases. Fig \ref{plot_sol_l0}(b) shows that the scaled scalar charge of phantom field increases monotonically from unity as $\beta$ increases. 

The metric component $g_{tt}$ is relevant with the the observer in the asymptotic region measure the redshift factor $z$, which describes the effect of gravitational redshift on a photon being emitted from a source in the wormhole spacetime \cite{Kleihaus:2020qwo},
\begin{equation}
 z = \frac{\lambda_\text{asym}}{\lambda_\text{emit}} = \frac{\sqrt{-g_{tt}(\infty)}}{\sqrt{-g_{tt}(0)}} -1 = \frac{1}{\sqrt{F_0(0)}}-1 \,,
\end{equation}
where $\lambda_\text{asym}$ is the wavelength measured by the observer and $\lambda_\text{emit}$ is the wavelength of photon in the wormhole. For simplicity, we consider the photon is emitted at the throat $\eta=0$. Fig. \ref{plot_sol_l0}(c) exhibits the profile for metric function $F_0$ is strictly increasing from $F_0(0)$ to the asymptotic value but the value $F_0(0)$ is strictly decreasing when $\beta$ increases. This gives rise the observer always measure the wavelength of photon being red--shifted as shown in Fig. \ref{plot_sol_l0}(d). Similarly, the profile of function $F_1$ in Fig. \ref{plot_sol_l0}(e) is strictly decreasing from its maximum value at $\eta=0$ to the asymptotic value. However, we find surprisingly that the wormholes still can possess double throat configuration, which we will discuss this in detail in the next paragraph. In Fig. \ref{plot_sol_l0}(f), the gauge field $K$ decays faster while the profile of gauge field $H$ doesn't vary too much when $\beta$ increases.

Let's turn our discussion to the geometry of these hairy wormholes. Fig. \ref{plot_prop}(a) shows that they possess only a single throat at $x=0$ within the range $0 \leq \beta < \beta_{\text{crit}}$ since $R''(0)>0$ in Fig. \ref{plot_prop}(b). Here $\beta_{\text{crit}}$ is the critical value of $\beta$ which approximately equal to 64.630655. When $\beta = \beta_{\text{crit}}$, we see that $R'(x_{\text{crit}})=R''(x_{\text{crit}})=0$ in Fig. \ref{plot_prop}(b), at this stage the wormhole simultaneously develops a throat at $x_{\text{th}}$ and an equator $x_{\text{eq}}$ at $x \approx 0.32237$, while $x=0$ is still maintaining as a throat. This means that the transition from single throat configuration to the double throat configuration can happen when $\beta \geq \beta_{\text{crit}}$, as shown in Fig. \ref{plot_prop}. In Fig. \ref{plot_prop}(a), the equator is always sandwiched between the throats $0 <  x_{\text{eq}} <x_{\text{th}}$. We also observe that the location of equator $x_{\text{eq}}$ moves toward $x=0$ and the location of another throat $x_{\text{th}}$ moves away from $x=0$, hence the equator and the throat at $x=0$ are very close. This gives rise to the circumferential radius of the equator is slightly larger than the circumferential radius of the throat at $x=0$ for very large values of $\beta$, as shown in Fig. \ref{plot_prop}(c). Here $R(0)$ increases monotonically as $\beta$ increases, which can be seen from Figs. \ref{plot_prop}(c) and (e).

When $\beta <  \beta_{\text{crit}}$, the surface gravity $\kappa$ vanishes for the wormholes with a single throat since $F'_0(0)=0$. However, another throat $\eta_{\text{th}}$ appears in between the equator and asymptotically flat region when $\beta=\beta_{\text{crit}}$, thus $\kappa$ assumes finite value, as shown in Fig. \ref{plot_prop}(d). When $\beta > \beta_{\text{crit}}$, $\kappa$ increases to a maximum value and then decreases for very large values of $\beta$.  Furthermore, we can visualize the structure of single throat and double throat in the embedding diagrams which are shown in Fig. \ref{plot_embed3d}.

In Fig. \ref{plot_prop}(f), we scale the NEC by $\alpha^2 (-T^t_t+T^\eta_\eta)$, so that we can compare the NEC of our wormholes with the Ellis wormhole. We found that the violation of NEC is the largest for the Ellis wormhole, particularly at the throat. When $\beta$ increases, the violation of NEC at the throat decreases and the surprise is the wormholes satisfy the NEC for large values of $\beta$. Note that the NEC vanishes at the asymptotic flat region for a very small value of $\beta$. However, the NEC assumes finite value at the asymptotic flat region when $\beta$ becomes very large, the reason is the first and third terms vanish but $H'$ doesn't vanish (can be seen from Fig. \ref{plot_sol_l0}(f)), thus the second term remain finite. In addition, the NEC at the infinity has been amplified due to the scaling factor $\alpha^2$.

\section{Conclusion and Outlook}

We have obtained the symmetric wormholes which supported by the phantom field in the Einstein--Yang--Mills--Higgs (EYMH) system in the Bogomol'nyi--Prasad--Sommerfield (BPS) limit. When we switch off the gravity, we obtain the probe limit which is the Yang--Mills--Higgs (YMH) field in the background of Ellis wormhole. In the presence of gravity, the wormholes possess the non--trivial non--abelian hair, thus the hairy wormholes solutions emerge from the Ellis wormhole where the wormholes gain the mass. The mass of wormholes and the scaled scalar charge of phantom field increase monotonically when the gravitational coupling constant increases. 

When the gravitational strength below a critical value, the wormholes only possess a single throat at the radial coordinate $\eta=0$, thus the corresponding surface gravity vanishes. When the gravitational strength equal to the critical value, an equator and another throat coexist simultaneously somewhere in the manifold, thus the transition from the single throat to the double throat can occur. The equator is sandwiched between the throat at $\eta=0$ and another throat. Therefore, the surface gravity of another throat assumes finite values. The circumferential radius of the throat at $\eta=0$ increases monotonically and still remains as the throat even in the strong gravitational field. The circumferential radius of the equator is slightly larger than the circumferential radius of the throat at $\eta=0$ because they are very close to each other in the large gravitational coupling. The violation of null energy condition is the largest for the Ellis wormhole, particularly at the throat. However, the violation of null energy condition decreases when the gravitational strength increases. Thus, the wormholes satisfy the null energy condition in the strong gravitational strength.

Since we only study the properties of hairy wormholes with the vanishing Higgs self--interaction, then the extension of this work would be natural to construct and investigate the properties of hairy wormholes with finite Higgs self--interaction. We have some preliminary results which show that the properties of hairy wormholes are different than the BPS case. With finite Higgs self--interaction, the mass of wormholes increases from zero but decreases very sharply when the gravitational strength reaches a critical value. The scaled scalar charge also increases very sharply at the critical value of gravitational strength. The results (in preparation) will be reported in elsewhere \cite{soonHiggsfinite}.

Concerning the stability issue of the wormholes, the wormholes solutions which are constructed by phantom field are generically unstable against the linear perturbation \cite{Gonzalez:2008wd,Gonzalez:2008xk,Torii:2013xba,Dzhunushaliev:2013lna,Dzhunushaliev:2014mza,Aringazin:2014rva}. Furthermore, the stability analysis has shown that the particle--like and hairy black hole solutions of EYMH system are unstable \cite{greene1993eluding,winstanley1995instability,mavromatos1996aspects}. Hence, we conjecture that the EYMH hairy wormholes are unstable as well because they will inherit the instabilities from the Ellis wormholes and behave qualitatively unstable as the compact objects in the EYMH system. It would be of interest to carry out a full linear stability analysis of hairy wormholes consistently by perturbing all the functions. However, the presence of YMH field could introduce the extra degree of freedom and cause the calculation of unstable modes to become non-trivial. Since the calculation of linear stability is tedious and requires a lot of effort, we leave this as an independent investigation. Nevertheless, the unstable modes disappear for sufficiently rapidly rotating Ellis wormholes in 5--dimensions with equal angular momenta \cite{Dzhunushaliev:2013jja}. Since the counterpart EYMH black holes can rotate, then it is interesting to construct the rotating EYMH wormholes which might be stable against the perturbations. 

%However, we leave the full analysis of the linear stability of hairy wormholes as a separate work because in order to perform linear stability in a consistent way, one has to perturb all the functions which are the non--Abelian gauge field $A_\mu$, two scalar fields $(\psi,H)$ and the background metric $g_{\mu \nu}$ up to the first order of the expansion in the radial perturbation, such that the contribution of other fields is not being suppressed. The radial perturbation yields a set of nonlinear ODEs which is the eigenvalue problem. The linear stability of wormholes is determined by the values of the modes (eigenvalue), in which indicates that the perturbation will either decay exponentially or increase exponentially with time. The presence of YMH field could introduce the extra degree of freedom, which makes the calculation of modes is not trivial, then one has to carefully study the asymptotic behaviour of the ODEs at the throat and infinity, such that one can calculate the stable or unstable modes by solving the ODEs numerically with the appropriate boundary conditions \cite{Blazquez-Salcedo:2018ipc}. Since the calculation is tedious and requires a lot of effort, we leave this as an independent investigation.

Since the static and regular EYMH solutions can also possess only axial symmetry and need not be spherically symmetric, their counterpart static black holes also can possess only axially symmetric horizon \cite{Hartmann:2000gx,Hartmann:2001ic} which is a counterexample to Israel's theorem. Therefore, as a first step to constructing the rotating wormholes in EYMH, we could consider constructing the static hairy wormhole solutions with a throat which is also axially symmetric.

%\begin{figure}[t!]
%\centering
%\mbox{
%(a)
% \includegraphics[angle =-90,scale=0.3]{plot_YMHw3a_r0=1_scaledmass}
%(b)
% \includegraphics[angle =-90,scale=0.3]{plot_YMHw3a_r0=1_D2scaled}
% }
%\caption{Properties of wormholes with finite Higgs self--interaction $\lambda$: a) The scaled mass $\mu/\alpha^2$ versus the scaled gravitational coupling constant $\alpha$; b) The logarithmic of scaled scalar charge $\log_{10}(\alpha^2 D^2)$ versus the scaled gravitational coupling constant $\alpha$}
%\label{plot_fullprop}
%\end{figure}

\section*{Acknowledgement}
XYC acknowledges the useful discussion with Jose Luis Bl\'azquez--Salcedo, Jutta Kunz, Yen Chin Ong and Eugen Radu. XYC thanks the National Research Foundation of Korea (grant no. 2018R1D1A1B07049126) for funding. 

\appendix

\bibliography{worm}

\newcommand{\noop}[1]{}
\begin{thebibliography}{85}
\expandafter\ifx\csname natexlab\endcsname\relax\def\natexlab#1{#1}\fi
\expandafter\ifx\csname bibnamefont\endcsname\relax
  \def\bibnamefont#1{#1}\fi
\expandafter\ifx\csname bibfnamefont\endcsname\relax
  \def\bibfnamefont#1{#1}\fi
\expandafter\ifx\csname citenamefont\endcsname\relax
  \def\citenamefont#1{#1}\fi
\expandafter\ifx\csname url\endcsname\relax
  \def\url#1{\texttt{#1}}\fi
\expandafter\ifx\csname urlprefix\endcsname\relax\def\urlprefix{URL }\fi
\providecommand{\bibinfo}[2]{#2}
\providecommand{\eprint}[2][]{\url{#2}}

\bibitem[{\citenamefont{t~Hooft}(1974)}]{t1974magnetic}
\bibinfo{author}{\bibfnamefont{G.}~\bibnamefont{t~Hooft}},
  \bibinfo{journal}{Nucl. Phys. B} \textbf{\bibinfo{volume}{79}},
  \bibinfo{pages}{276} (\bibinfo{year}{1974}).

\bibitem[{\citenamefont{Polyakov}(1974)}]{polyakov1974particle}
\bibinfo{author}{\bibfnamefont{A.~M.} \bibnamefont{Polyakov}},
  \bibinfo{journal}{JETP Letters} \textbf{\bibinfo{volume}{20}},
  \bibinfo{pages}{194} (\bibinfo{year}{1974}).

\bibitem[{\citenamefont{Rebbi and Rossi}(1980)}]{rebbi1980multimonopole}
\bibinfo{author}{\bibfnamefont{C.}~\bibnamefont{Rebbi}} \bibnamefont{and}
  \bibinfo{author}{\bibfnamefont{P.}~\bibnamefont{Rossi}},
  \bibinfo{journal}{Phys. Rev. D} \textbf{\bibinfo{volume}{22}},
  \bibinfo{pages}{2010} (\bibinfo{year}{1980}).

\bibitem[{\citenamefont{Ward}(1981)}]{ward1981yang}
\bibinfo{author}{\bibfnamefont{R.}~\bibnamefont{Ward}},
  \bibinfo{journal}{Commun. Math. Phys.} \textbf{\bibinfo{volume}{79}},
  \bibinfo{pages}{317} (\bibinfo{year}{1981}).

\bibitem[{\citenamefont{Forg{\'a}cs et~al.}(1981)\citenamefont{Forg{\'a}cs,
  Horv{\'a}th, and Palla}}]{forgacs1981exact}
\bibinfo{author}{\bibfnamefont{P.}~\bibnamefont{Forg{\'a}cs}},
  \bibinfo{author}{\bibfnamefont{Z.}~\bibnamefont{Horv{\'a}th}},
  \bibnamefont{and} \bibinfo{author}{\bibfnamefont{L.}~\bibnamefont{Palla}},
  \bibinfo{journal}{Phys. Lett. B} \textbf{\bibinfo{volume}{99}},
  \bibinfo{pages}{232} (\bibinfo{year}{1981}).

\bibitem[{\citenamefont{Prasad and Rossi}(1981)}]{prasad1981construction}
\bibinfo{author}{\bibfnamefont{M.}~\bibnamefont{Prasad}} \bibnamefont{and}
  \bibinfo{author}{\bibfnamefont{P.}~\bibnamefont{Rossi}},
  \bibinfo{journal}{Phys. Rev. D} \textbf{\bibinfo{volume}{24}},
  \bibinfo{pages}{2182} (\bibinfo{year}{1981}).

\bibitem[{\citenamefont{Prasad and Sommerfield}(1975)}]{prasad1975exact}
\bibinfo{author}{\bibfnamefont{M.}~\bibnamefont{Prasad}} \bibnamefont{and}
  \bibinfo{author}{\bibfnamefont{C.~M.} \bibnamefont{Sommerfield}},
  \bibinfo{journal}{Phys. Rev. Lett.} \textbf{\bibinfo{volume}{35}},
  \bibinfo{pages}{760} (\bibinfo{year}{1975}).

\bibitem[{\citenamefont{Bogomol'Nyi}(1976)}]{bogomol1976stability}
\bibinfo{author}{\bibfnamefont{E.}~\bibnamefont{Bogomol'Nyi}},
  \bibinfo{journal}{Sov. J. Nucl. Phys.} \textbf{\bibinfo{volume}{24}}
  (\bibinfo{year}{1976}).

\bibitem[{\citenamefont{Kleihaus et~al.}(1998)\citenamefont{Kleihaus, Kunz, and
  Tchrakian}}]{kleihaus1998interaction}
\bibinfo{author}{\bibfnamefont{B.}~\bibnamefont{Kleihaus}},
  \bibinfo{author}{\bibfnamefont{J.}~\bibnamefont{Kunz}}, \bibnamefont{and}
  \bibinfo{author}{\bibfnamefont{D.}~\bibnamefont{Tchrakian}},
  \bibinfo{journal}{Modern Physics Letters A} \textbf{\bibinfo{volume}{13}},
  \bibinfo{pages}{2523} (\bibinfo{year}{1998}).

\bibitem[{\citenamefont{Kleihaus and Kunz}(1999)}]{kleihaus1999monopole}
\bibinfo{author}{\bibfnamefont{B.}~\bibnamefont{Kleihaus}} \bibnamefont{and}
  \bibinfo{author}{\bibfnamefont{J.}~\bibnamefont{Kunz}},
  \bibinfo{journal}{Phys. Rev. D} \textbf{\bibinfo{volume}{61}},
  \bibinfo{pages}{025003} (\bibinfo{year}{1999}).

\bibitem[{\citenamefont{Kleihaus et~al.}(2003)\citenamefont{Kleihaus, Kunz, and
  Shnir}}]{kleihaus2003monopole}
\bibinfo{author}{\bibfnamefont{B.}~\bibnamefont{Kleihaus}},
  \bibinfo{author}{\bibfnamefont{J.}~\bibnamefont{Kunz}}, \bibnamefont{and}
  \bibinfo{author}{\bibfnamefont{Y.}~\bibnamefont{Shnir}},
  \bibinfo{journal}{Phys. Lett. B} \textbf{\bibinfo{volume}{570}},
  \bibinfo{pages}{237} (\bibinfo{year}{2003}).

\bibitem[{\citenamefont{Kleihaus
  et~al.}(2004{\natexlab{a}})\citenamefont{Kleihaus, Kunz, and
  Shnir}}]{kleihaus2004monopole}
\bibinfo{author}{\bibfnamefont{B.}~\bibnamefont{Kleihaus}},
  \bibinfo{author}{\bibfnamefont{J.}~\bibnamefont{Kunz}}, \bibnamefont{and}
  \bibinfo{author}{\bibfnamefont{Y.}~\bibnamefont{Shnir}},
  \bibinfo{journal}{Physical Review D} \textbf{\bibinfo{volume}{70}},
  \bibinfo{pages}{065010} (\bibinfo{year}{2004}{\natexlab{a}}).

\bibitem[{\citenamefont{Breitenlohner et~al.}(1992)\citenamefont{Breitenlohner,
  Forgacs, and Maison}}]{Breitenlohner:1991aa}
\bibinfo{author}{\bibfnamefont{P.}~\bibnamefont{Breitenlohner}},
  \bibinfo{author}{\bibfnamefont{P.}~\bibnamefont{Forgacs}}, \bibnamefont{and}
  \bibinfo{author}{\bibfnamefont{D.}~\bibnamefont{Maison}},
  \bibinfo{journal}{Nucl. Phys. B} \textbf{\bibinfo{volume}{383}},
  \bibinfo{pages}{357} (\bibinfo{year}{1992}).

\bibitem[{\citenamefont{Lee et~al.}(1992)\citenamefont{Lee, Nair, and
  Weinberg}}]{lee1992black}
\bibinfo{author}{\bibfnamefont{K.}~\bibnamefont{Lee}},
  \bibinfo{author}{\bibfnamefont{V.}~\bibnamefont{Nair}}, \bibnamefont{and}
  \bibinfo{author}{\bibfnamefont{E.~J.} \bibnamefont{Weinberg}},
  \bibinfo{journal}{Phys. Rev. D} \textbf{\bibinfo{volume}{45}},
  \bibinfo{pages}{2751} (\bibinfo{year}{1992}).

\bibitem[{\citenamefont{Breitenlohner et~al.}(1995)\citenamefont{Breitenlohner,
  Forgacs, and Maison}}]{Breitenlohner:1994di}
\bibinfo{author}{\bibfnamefont{P.}~\bibnamefont{Breitenlohner}},
  \bibinfo{author}{\bibfnamefont{P.}~\bibnamefont{Forgacs}}, \bibnamefont{and}
  \bibinfo{author}{\bibfnamefont{D.}~\bibnamefont{Maison}},
  \bibinfo{journal}{Nucl. Phys. B} \textbf{\bibinfo{volume}{442}},
  \bibinfo{pages}{126} (\bibinfo{year}{1995}), \eprint{gr-qc/9412039}.

\bibitem[{\citenamefont{Brihaye et~al.}(1998)\citenamefont{Brihaye, Hartmann,
  and Kunz}}]{Brihaye:1998cm}
\bibinfo{author}{\bibfnamefont{Y.}~\bibnamefont{Brihaye}},
  \bibinfo{author}{\bibfnamefont{B.}~\bibnamefont{Hartmann}}, \bibnamefont{and}
  \bibinfo{author}{\bibfnamefont{J.}~\bibnamefont{Kunz}},
  \bibinfo{journal}{Phys. Lett. B} \textbf{\bibinfo{volume}{441}},
  \bibinfo{pages}{77} (\bibinfo{year}{1998}).

\bibitem[{\citenamefont{Lue and Weinberg}(1999)}]{Lue:1999zp}
\bibinfo{author}{\bibfnamefont{A.}~\bibnamefont{Lue}} \bibnamefont{and}
  \bibinfo{author}{\bibfnamefont{E.~J.} \bibnamefont{Weinberg}},
  \bibinfo{journal}{Phys. Rev. D} \textbf{\bibinfo{volume}{60}},
  \bibinfo{pages}{084025} (\bibinfo{year}{1999}), \eprint{hep-th/9905223}.

\bibitem[{\citenamefont{Brihaye et~al.}(1999)\citenamefont{Brihaye, Hartmann,
  Kunz, and Tell}}]{Brihaye:1999nn}
\bibinfo{author}{\bibfnamefont{Y.}~\bibnamefont{Brihaye}},
  \bibinfo{author}{\bibfnamefont{B.}~\bibnamefont{Hartmann}},
  \bibinfo{author}{\bibfnamefont{J.}~\bibnamefont{Kunz}}, \bibnamefont{and}
  \bibinfo{author}{\bibfnamefont{N.}~\bibnamefont{Tell}},
  \bibinfo{journal}{Phys. Rev. D} \textbf{\bibinfo{volume}{60}},
  \bibinfo{pages}{104016} (\bibinfo{year}{1999}), \eprint{hep-th/9904065}.

\bibitem[{\citenamefont{Brihaye
  et~al.}(2000{\natexlab{a}})\citenamefont{Brihaye, Hartmann, and
  Kunz}}]{Brihaye:1999kt}
\bibinfo{author}{\bibfnamefont{Y.}~\bibnamefont{Brihaye}},
  \bibinfo{author}{\bibfnamefont{B.}~\bibnamefont{Hartmann}}, \bibnamefont{and}
  \bibinfo{author}{\bibfnamefont{J.}~\bibnamefont{Kunz}},
  \bibinfo{journal}{Phys. Rev. D} \textbf{\bibinfo{volume}{62}},
  \bibinfo{pages}{044008} (\bibinfo{year}{2000}{\natexlab{a}}),
  \eprint{hep-th/9911148}.

\bibitem[{\citenamefont{Brihaye
  et~al.}(2000{\natexlab{b}})\citenamefont{Brihaye, Grard, and
  Hoorelbeke}}]{Brihaye:1999zt}
\bibinfo{author}{\bibfnamefont{Y.}~\bibnamefont{Brihaye}},
  \bibinfo{author}{\bibfnamefont{F.}~\bibnamefont{Grard}}, \bibnamefont{and}
  \bibinfo{author}{\bibfnamefont{S.}~\bibnamefont{Hoorelbeke}},
  \bibinfo{journal}{Phys. Rev. D} \textbf{\bibinfo{volume}{62}},
  \bibinfo{pages}{044013} (\bibinfo{year}{2000}{\natexlab{b}}),
  \eprint{hep-th/9912023}.

\bibitem[{\citenamefont{Brihaye and Hartmann}(2002)}]{Brihaye:2002pc}
\bibinfo{author}{\bibfnamefont{Y.}~\bibnamefont{Brihaye}} \bibnamefont{and}
  \bibinfo{author}{\bibfnamefont{B.}~\bibnamefont{Hartmann}},
  \bibinfo{journal}{Phys. Rev. D} \textbf{\bibinfo{volume}{66}},
  \bibinfo{pages}{064018} (\bibinfo{year}{2002}), \eprint{hep-th/0206004}.

\bibitem[{\citenamefont{Hartmann
  et~al.}(2001{\natexlab{a}})\citenamefont{Hartmann, Kleihaus, and
  Kunz}}]{Hartmann:2000gx}
\bibinfo{author}{\bibfnamefont{B.}~\bibnamefont{Hartmann}},
  \bibinfo{author}{\bibfnamefont{B.}~\bibnamefont{Kleihaus}}, \bibnamefont{and}
  \bibinfo{author}{\bibfnamefont{J.}~\bibnamefont{Kunz}},
  \bibinfo{journal}{Phys. Rev. Lett.} \textbf{\bibinfo{volume}{86}},
  \bibinfo{pages}{1422} (\bibinfo{year}{2001}{\natexlab{a}}),
  \eprint{hep-th/0009195}.

\bibitem[{\citenamefont{Hartmann
  et~al.}(2001{\natexlab{b}})\citenamefont{Hartmann, Kleihaus, and
  Kunz}}]{Hartmann:2001ic}
\bibinfo{author}{\bibfnamefont{B.}~\bibnamefont{Hartmann}},
  \bibinfo{author}{\bibfnamefont{B.}~\bibnamefont{Kleihaus}}, \bibnamefont{and}
  \bibinfo{author}{\bibfnamefont{J.}~\bibnamefont{Kunz}},
  \bibinfo{journal}{Phys. Rev. D} \textbf{\bibinfo{volume}{65}},
  \bibinfo{pages}{024027} (\bibinfo{year}{2001}{\natexlab{b}}).

\bibitem[{\citenamefont{Kleihaus and
  Kunz}(2000{\natexlab{a}})}]{Kleihaus:2000hx}
\bibinfo{author}{\bibfnamefont{B.}~\bibnamefont{Kleihaus}} \bibnamefont{and}
  \bibinfo{author}{\bibfnamefont{J.}~\bibnamefont{Kunz}},
  \bibinfo{journal}{Phys. Rev. Lett.} \textbf{\bibinfo{volume}{85}},
  \bibinfo{pages}{2430} (\bibinfo{year}{2000}{\natexlab{a}}),
  \eprint{hep-th/0006148}.

\bibitem[{\citenamefont{Kleihaus and
  Kunz}(2000{\natexlab{b}})}]{Kleihaus:2000kv}
\bibinfo{author}{\bibfnamefont{B.}~\bibnamefont{Kleihaus}} \bibnamefont{and}
  \bibinfo{author}{\bibfnamefont{J.}~\bibnamefont{Kunz}},
  \bibinfo{journal}{Phys. Lett. B} \textbf{\bibinfo{volume}{494}},
  \bibinfo{pages}{130} (\bibinfo{year}{2000}{\natexlab{b}}),
  \eprint{hep-th/0008034}.

\bibitem[{\citenamefont{Van~der Bij and Radu}(2002)}]{VanderBij:2001nm}
\bibinfo{author}{\bibfnamefont{J.}~\bibnamefont{Van~der Bij}} \bibnamefont{and}
  \bibinfo{author}{\bibfnamefont{E.}~\bibnamefont{Radu}},
  \bibinfo{journal}{Int. J. Mod. Phys. A} \textbf{\bibinfo{volume}{17}},
  \bibinfo{pages}{1477} (\bibinfo{year}{2002}), \eprint{gr-qc/0111046}.

\bibitem[{\citenamefont{Paturyan and Tchrakian}(2004)}]{Paturyan:2003hz}
\bibinfo{author}{\bibfnamefont{V.}~\bibnamefont{Paturyan}} \bibnamefont{and}
  \bibinfo{author}{\bibfnamefont{D.}~\bibnamefont{Tchrakian}},
  \bibinfo{journal}{J. Math. Phys.} \textbf{\bibinfo{volume}{45}},
  \bibinfo{pages}{302} (\bibinfo{year}{2004}), \eprint{hep-th/0306160}.

\bibitem[{\citenamefont{Paturyan et~al.}(2005)\citenamefont{Paturyan, Radu, and
  Tchrakian}}]{Paturyan:2004ps}
\bibinfo{author}{\bibfnamefont{V.}~\bibnamefont{Paturyan}},
  \bibinfo{author}{\bibfnamefont{E.}~\bibnamefont{Radu}}, \bibnamefont{and}
  \bibinfo{author}{\bibfnamefont{D.}~\bibnamefont{Tchrakian}},
  \bibinfo{journal}{Phys. Lett. B} \textbf{\bibinfo{volume}{609}},
  \bibinfo{pages}{360} (\bibinfo{year}{2005}), \eprint{hep-th/0412011}.

\bibitem[{\citenamefont{Kleihaus
  et~al.}(2004{\natexlab{b}})\citenamefont{Kleihaus, Kunz, and
  Navarro-Lerida}}]{Kleihaus:2004gm}
\bibinfo{author}{\bibfnamefont{B.}~\bibnamefont{Kleihaus}},
  \bibinfo{author}{\bibfnamefont{J.}~\bibnamefont{Kunz}}, \bibnamefont{and}
  \bibinfo{author}{\bibfnamefont{F.}~\bibnamefont{Navarro-Lerida}},
  \bibinfo{journal}{Phys. Lett. B} \textbf{\bibinfo{volume}{599}},
  \bibinfo{pages}{294} (\bibinfo{year}{2004}{\natexlab{b}}),
  \eprint{gr-qc/0406094}.

\bibitem[{\citenamefont{Kleihaus
  et~al.}(2005{\natexlab{a}})\citenamefont{Kleihaus, Kunz, and
  Shnir}}]{Kleihaus:2004fh}
\bibinfo{author}{\bibfnamefont{B.}~\bibnamefont{Kleihaus}},
  \bibinfo{author}{\bibfnamefont{J.}~\bibnamefont{Kunz}}, \bibnamefont{and}
  \bibinfo{author}{\bibfnamefont{Y.}~\bibnamefont{Shnir}},
  \bibinfo{journal}{Phys. Rev. D} \textbf{\bibinfo{volume}{71}},
  \bibinfo{pages}{024013} (\bibinfo{year}{2005}{\natexlab{a}}),
  \eprint{gr-qc/0411106}.

\bibitem[{\citenamefont{Kleihaus
  et~al.}(2005{\natexlab{b}})\citenamefont{Kleihaus, Kunz, and
  Neemann}}]{Kleihaus:2005fs}
\bibinfo{author}{\bibfnamefont{B.}~\bibnamefont{Kleihaus}},
  \bibinfo{author}{\bibfnamefont{J.}~\bibnamefont{Kunz}}, \bibnamefont{and}
  \bibinfo{author}{\bibfnamefont{U.}~\bibnamefont{Neemann}},
  \bibinfo{journal}{Phys. Lett. B} \textbf{\bibinfo{volume}{623}},
  \bibinfo{pages}{171} (\bibinfo{year}{2005}{\natexlab{b}}),
  \eprint{gr-qc/0507047}.

\bibitem[{\citenamefont{Hauser et~al.}(2014)\citenamefont{Hauser, Ibadov,
  Kleihaus, and Kunz}}]{hauser2014hairy}
\bibinfo{author}{\bibfnamefont{O.}~\bibnamefont{Hauser}},
  \bibinfo{author}{\bibfnamefont{R.}~\bibnamefont{Ibadov}},
  \bibinfo{author}{\bibfnamefont{B.}~\bibnamefont{Kleihaus}}, \bibnamefont{and}
  \bibinfo{author}{\bibfnamefont{J.}~\bibnamefont{Kunz}},
  \bibinfo{journal}{Phys. Rev. D} \textbf{\bibinfo{volume}{89}},
  \bibinfo{pages}{064010} (\bibinfo{year}{2014}).

\bibitem[{\citenamefont{Bartnik and McKinnon}(1988)}]{bartnik1988particlelike}
\bibinfo{author}{\bibfnamefont{R.}~\bibnamefont{Bartnik}} \bibnamefont{and}
  \bibinfo{author}{\bibfnamefont{J.}~\bibnamefont{McKinnon}},
  \bibinfo{journal}{Phys. Rev. Lett.} \textbf{\bibinfo{volume}{61}},
  \bibinfo{pages}{141} (\bibinfo{year}{1988}).

\bibitem[{\citenamefont{Caldwell}(2002)}]{Caldwell:1999ew}
\bibinfo{author}{\bibfnamefont{R.~R.} \bibnamefont{Caldwell}},
  \bibinfo{journal}{Phys. Lett. B} \textbf{\bibinfo{volume}{545}},
  \bibinfo{pages}{23} (\bibinfo{year}{2002}).

\bibitem[{\citenamefont{Carroll et~al.}(2003)\citenamefont{Carroll, Hoffman,
  and Trodden}}]{Carroll:2003st}
\bibinfo{author}{\bibfnamefont{S.~M.} \bibnamefont{Carroll}},
  \bibinfo{author}{\bibfnamefont{M.}~\bibnamefont{Hoffman}}, \bibnamefont{and}
  \bibinfo{author}{\bibfnamefont{M.}~\bibnamefont{Trodden}},
  \bibinfo{journal}{Phys. Rev. D} \textbf{\bibinfo{volume}{68}},
  \bibinfo{pages}{023509} (\bibinfo{year}{2003}).

\bibitem[{\citenamefont{Gibbons}(2003)}]{Gibbons:2003yj}
\bibinfo{author}{\bibfnamefont{G.}~\bibnamefont{Gibbons}},
  \bibinfo{journal}{arXiv preprint hep-th/0302199}  (\bibinfo{year}{2003}).

\bibitem[{\citenamefont{Hannestad}(2006)}]{Hannestad:2005fg}
\bibinfo{author}{\bibfnamefont{S.}~\bibnamefont{Hannestad}},
  \bibinfo{journal}{Int. J. Mod. Phys. A} \textbf{\bibinfo{volume}{21}},
  \bibinfo{pages}{1938} (\bibinfo{year}{2006}).

\bibitem[{\citenamefont{Bronnikov and Fabris}(2006)}]{Bronnikov:2005gm}
\bibinfo{author}{\bibfnamefont{K.~A.} \bibnamefont{Bronnikov}}
  \bibnamefont{and} \bibinfo{author}{\bibfnamefont{J.~C.}
  \bibnamefont{Fabris}}, \bibinfo{journal}{Phys. Rev. Lett.}
  \textbf{\bibinfo{volume}{96}}, \bibinfo{pages}{251101}
  (\bibinfo{year}{2006}).

\bibitem[{\citenamefont{Chen et~al.}(2016)\citenamefont{Chen, Wang, and
  Jing}}]{Chen:2016yey}
\bibinfo{author}{\bibfnamefont{S.}~\bibnamefont{Chen}},
  \bibinfo{author}{\bibfnamefont{M.}~\bibnamefont{Wang}}, \bibnamefont{and}
  \bibinfo{author}{\bibfnamefont{J.}~\bibnamefont{Jing}},
  \bibinfo{journal}{Class. Quantum Gravity} \textbf{\bibinfo{volume}{33}},
  \bibinfo{pages}{195002} (\bibinfo{year}{2016}).

\bibitem[{\citenamefont{Kleihaus et~al.}(2019)\citenamefont{Kleihaus, Kunz, and
  Radu}}]{Kleihaus:2019wck}
\bibinfo{author}{\bibfnamefont{B.}~\bibnamefont{Kleihaus}},
  \bibinfo{author}{\bibfnamefont{J.}~\bibnamefont{Kunz}}, \bibnamefont{and}
  \bibinfo{author}{\bibfnamefont{E.}~\bibnamefont{Radu}},
  \bibinfo{journal}{Phys. Lett. B} \textbf{\bibinfo{volume}{797}},
  \bibinfo{pages}{134892} (\bibinfo{year}{2019}).

\bibitem[{\citenamefont{Dzhunushaliev et~al.}(2008)\citenamefont{Dzhunushaliev,
  Folomeev, Myrzakulov, and Singleton}}]{Dzhunushaliev:2008bq}
\bibinfo{author}{\bibfnamefont{V.}~\bibnamefont{Dzhunushaliev}},
  \bibinfo{author}{\bibfnamefont{V.}~\bibnamefont{Folomeev}},
  \bibinfo{author}{\bibfnamefont{R.}~\bibnamefont{Myrzakulov}},
  \bibnamefont{and}
  \bibinfo{author}{\bibfnamefont{D.}~\bibnamefont{Singleton}},
  \bibinfo{journal}{J. High Energy Phys.} \textbf{\bibinfo{volume}{2008}},
  \bibinfo{pages}{094} (\bibinfo{year}{2008}).

\bibitem[{\citenamefont{Ellis}(1973)}]{Ellis:1973yv}
\bibinfo{author}{\bibfnamefont{H.~G.} \bibnamefont{Ellis}},
  \bibinfo{journal}{J. Math. Phys.} \textbf{\bibinfo{volume}{14}},
  \bibinfo{pages}{104} (\bibinfo{year}{1973}).

\bibitem[{\citenamefont{Ellis}(1979)}]{Ellis:1979bh}
\bibinfo{author}{\bibfnamefont{H.~G.} \bibnamefont{Ellis}},
  \bibinfo{journal}{Gen. Relativ. Gravit.} \textbf{\bibinfo{volume}{10}},
  \bibinfo{pages}{105} (\bibinfo{year}{1979}).

\bibitem[{\citenamefont{Bronnikov}(1973)}]{Bronnikov:1973fh}
\bibinfo{author}{\bibfnamefont{K.~A.} \bibnamefont{Bronnikov}},
  \bibinfo{journal}{Acta. Phys. Pol} p.~\bibinfo{pages}{B4}
  (\bibinfo{year}{1973}).

\bibitem[{\citenamefont{Visser}(1995)}]{Visser:1995cc}
\bibinfo{author}{\bibfnamefont{M.}~\bibnamefont{Visser}},
  \emph{\bibinfo{title}{Lorentzian Wormholes. From Einstein to Hawking}}
  (\bibinfo{publisher}{Woodbury, USA: AIP}, \bibinfo{year}{1995}).

\bibitem[{\citenamefont{Kanti et~al.}(2011)\citenamefont{Kanti, Kleihaus, and
  Kunz}}]{Kanti:2011jz}
\bibinfo{author}{\bibfnamefont{P.}~\bibnamefont{Kanti}},
  \bibinfo{author}{\bibfnamefont{B.}~\bibnamefont{Kleihaus}}, \bibnamefont{and}
  \bibinfo{author}{\bibfnamefont{J.}~\bibnamefont{Kunz}},
  \bibinfo{journal}{Phys. Rev. Lett.} \textbf{\bibinfo{volume}{107}},
  \bibinfo{pages}{271101} (\bibinfo{year}{2011}).

\bibitem[{\citenamefont{Kanti et~al.}(2012)\citenamefont{Kanti, Kleihaus, and
  Kunz}}]{Kanti:2011yv}
\bibinfo{author}{\bibfnamefont{P.}~\bibnamefont{Kanti}},
  \bibinfo{author}{\bibfnamefont{B.}~\bibnamefont{Kleihaus}}, \bibnamefont{and}
  \bibinfo{author}{\bibfnamefont{J.}~\bibnamefont{Kunz}},
  \bibinfo{journal}{Phys. Rev. D} \textbf{\bibinfo{volume}{85}},
  \bibinfo{pages}{044007} (\bibinfo{year}{2012}).

\bibitem[{\citenamefont{Antoniou et~al.}(2020)\citenamefont{Antoniou,
  Bakopoulos, Kanti, Kleihaus, and Kunz}}]{Antoniou:2019awm}
\bibinfo{author}{\bibfnamefont{G.}~\bibnamefont{Antoniou}},
  \bibinfo{author}{\bibfnamefont{A.}~\bibnamefont{Bakopoulos}},
  \bibinfo{author}{\bibfnamefont{P.}~\bibnamefont{Kanti}},
  \bibinfo{author}{\bibfnamefont{B.}~\bibnamefont{Kleihaus}}, \bibnamefont{and}
  \bibinfo{author}{\bibfnamefont{J.}~\bibnamefont{Kunz}},
  \bibinfo{journal}{Phys. Rev. D} \textbf{\bibinfo{volume}{101}},
  \bibinfo{pages}{024033} (\bibinfo{year}{2020}), \eprint{1904.13091}.

\bibitem[{\citenamefont{Gonzalez
  et~al.}(2009{\natexlab{a}})\citenamefont{Gonzalez, Guzman, and
  Sarbach}}]{Gonzalez:2008wd}
\bibinfo{author}{\bibfnamefont{J.}~\bibnamefont{Gonzalez}},
  \bibinfo{author}{\bibfnamefont{F.}~\bibnamefont{Guzman}}, \bibnamefont{and}
  \bibinfo{author}{\bibfnamefont{O.}~\bibnamefont{Sarbach}},
  \bibinfo{journal}{Class. Quant. Grav.} \textbf{\bibinfo{volume}{26}},
  \bibinfo{pages}{015010} (\bibinfo{year}{2009}{\natexlab{a}}),
  \eprint{0806.0608}.

\bibitem[{\citenamefont{Gonzalez
  et~al.}(2009{\natexlab{b}})\citenamefont{Gonzalez, Guzman, and
  Sarbach}}]{Gonzalez:2008xk}
\bibinfo{author}{\bibfnamefont{J.}~\bibnamefont{Gonzalez}},
  \bibinfo{author}{\bibfnamefont{F.}~\bibnamefont{Guzman}}, \bibnamefont{and}
  \bibinfo{author}{\bibfnamefont{O.}~\bibnamefont{Sarbach}},
  \bibinfo{journal}{Class. Quant. Grav.} \textbf{\bibinfo{volume}{26}},
  \bibinfo{pages}{015011} (\bibinfo{year}{2009}{\natexlab{b}}),
  \eprint{0806.1370}.

\bibitem[{\citenamefont{Torii and Shinkai}(2013)}]{Torii:2013xba}
\bibinfo{author}{\bibfnamefont{T.}~\bibnamefont{Torii}} \bibnamefont{and}
  \bibinfo{author}{\bibfnamefont{H.-a.} \bibnamefont{Shinkai}},
  \bibinfo{journal}{Phys. Rev. D} \textbf{\bibinfo{volume}{88}},
  \bibinfo{pages}{064027} (\bibinfo{year}{2013}).

\bibitem[{\citenamefont{Kashargin and
  Sushkov}(2008{\natexlab{a}})}]{Kashargin:2007mm}
\bibinfo{author}{\bibfnamefont{P.}~\bibnamefont{Kashargin}} \bibnamefont{and}
  \bibinfo{author}{\bibfnamefont{S.}~\bibnamefont{Sushkov}},
  \bibinfo{journal}{Gravit. Cosmol.} \textbf{\bibinfo{volume}{14}},
  \bibinfo{pages}{80} (\bibinfo{year}{2008}{\natexlab{a}}).

\bibitem[{\citenamefont{Kashargin and
  Sushkov}(2008{\natexlab{b}})}]{Kashargin:2008pk}
\bibinfo{author}{\bibfnamefont{P.}~\bibnamefont{Kashargin}} \bibnamefont{and}
  \bibinfo{author}{\bibfnamefont{S.}~\bibnamefont{Sushkov}},
  \bibinfo{journal}{Phys. Rev. D} \textbf{\bibinfo{volume}{78}},
  \bibinfo{pages}{064071} (\bibinfo{year}{2008}{\natexlab{b}}).

\bibitem[{\citenamefont{Kleihaus and Kunz}(2014)}]{Kleihaus:2014dla}
\bibinfo{author}{\bibfnamefont{B.}~\bibnamefont{Kleihaus}} \bibnamefont{and}
  \bibinfo{author}{\bibfnamefont{J.}~\bibnamefont{Kunz}},
  \bibinfo{journal}{Phys. Rev. D} \textbf{\bibinfo{volume}{90}},
  \bibinfo{pages}{121503} (\bibinfo{year}{2014}).

\bibitem[{\citenamefont{Chew et~al.}(2016)\citenamefont{Chew, Kleihaus, and
  Kunz}}]{Chew:2016epf}
\bibinfo{author}{\bibfnamefont{X.~Y.} \bibnamefont{Chew}},
  \bibinfo{author}{\bibfnamefont{B.}~\bibnamefont{Kleihaus}}, \bibnamefont{and}
  \bibinfo{author}{\bibfnamefont{J.}~\bibnamefont{Kunz}},
  \bibinfo{journal}{Phys. Rev. D} \textbf{\bibinfo{volume}{94}},
  \bibinfo{pages}{104031} (\bibinfo{year}{2016}).

\bibitem[{\citenamefont{Dzhunushaliev
  et~al.}(2013{\natexlab{a}})\citenamefont{Dzhunushaliev, Folomeev, Kleihaus,
  Kunz, and Radu}}]{Dzhunushaliev:2013jja}
\bibinfo{author}{\bibfnamefont{V.}~\bibnamefont{Dzhunushaliev}},
  \bibinfo{author}{\bibfnamefont{V.}~\bibnamefont{Folomeev}},
  \bibinfo{author}{\bibfnamefont{B.}~\bibnamefont{Kleihaus}},
  \bibinfo{author}{\bibfnamefont{J.}~\bibnamefont{Kunz}}, \bibnamefont{and}
  \bibinfo{author}{\bibfnamefont{E.}~\bibnamefont{Radu}},
  \bibinfo{journal}{Phys. Rev. D} \textbf{\bibinfo{volume}{88}},
  \bibinfo{pages}{124028} (\bibinfo{year}{2013}{\natexlab{a}}).

\bibitem[{\citenamefont{Chew et~al.}(2018)\citenamefont{Chew, Kleihaus, and
  Kunz}}]{Chew:2018vjp}
\bibinfo{author}{\bibfnamefont{X.~Y.} \bibnamefont{Chew}},
  \bibinfo{author}{\bibfnamefont{B.}~\bibnamefont{Kleihaus}}, \bibnamefont{and}
  \bibinfo{author}{\bibfnamefont{J.}~\bibnamefont{Kunz}},
  \bibinfo{journal}{Phys. Rev. D} \textbf{\bibinfo{volume}{97}},
  \bibinfo{pages}{064026} (\bibinfo{year}{2018}).

\bibitem[{\citenamefont{Nedkova et~al.}(2013)\citenamefont{Nedkova, Tinchev,
  and Yazadjiev}}]{Nedkova:2013msa}
\bibinfo{author}{\bibfnamefont{P.~G.} \bibnamefont{Nedkova}},
  \bibinfo{author}{\bibfnamefont{V.~K.} \bibnamefont{Tinchev}},
  \bibnamefont{and} \bibinfo{author}{\bibfnamefont{S.~S.}
  \bibnamefont{Yazadjiev}}, \bibinfo{journal}{Phys. Rev. D}
  \textbf{\bibinfo{volume}{88}}, \bibinfo{pages}{124019}
  (\bibinfo{year}{2013}).

\bibitem[{\citenamefont{Gyulchev et~al.}(2018)\citenamefont{Gyulchev, Nedkova,
  Tinchev, and Yazadjiev}}]{Gyulchev:2018fmd}
\bibinfo{author}{\bibfnamefont{G.}~\bibnamefont{Gyulchev}},
  \bibinfo{author}{\bibfnamefont{P.}~\bibnamefont{Nedkova}},
  \bibinfo{author}{\bibfnamefont{V.}~\bibnamefont{Tinchev}}, \bibnamefont{and}
  \bibinfo{author}{\bibfnamefont{S.}~\bibnamefont{Yazadjiev}},
  \bibinfo{journal}{Eur. Phys. J. C} \textbf{\bibinfo{volume}{78}},
  \bibinfo{pages}{544} (\bibinfo{year}{2018}).

\bibitem[{\citenamefont{Amir et~al.}(2019)\citenamefont{Amir, Banerjee, and
  Maharaj}}]{Amir:2018szm}
\bibinfo{author}{\bibfnamefont{M.}~\bibnamefont{Amir}},
  \bibinfo{author}{\bibfnamefont{A.}~\bibnamefont{Banerjee}}, \bibnamefont{and}
  \bibinfo{author}{\bibfnamefont{S.~D.} \bibnamefont{Maharaj}},
  \bibinfo{journal}{Ann. Phys.} \textbf{\bibinfo{volume}{400}},
  \bibinfo{pages}{198} (\bibinfo{year}{2019}).

\bibitem[{\citenamefont{Abe}(2010)}]{Abe:2010ap}
\bibinfo{author}{\bibfnamefont{F.}~\bibnamefont{Abe}},
  \bibinfo{journal}{Astrophys. J.} \textbf{\bibinfo{volume}{725}},
  \bibinfo{pages}{787} (\bibinfo{year}{2010}).

\bibitem[{\citenamefont{Toki et~al.}(2011)\citenamefont{Toki, Kitamura, Asada,
  and Abe}}]{Toki:2011zu}
\bibinfo{author}{\bibfnamefont{Y.}~\bibnamefont{Toki}},
  \bibinfo{author}{\bibfnamefont{T.}~\bibnamefont{Kitamura}},
  \bibinfo{author}{\bibfnamefont{H.}~\bibnamefont{Asada}}, \bibnamefont{and}
  \bibinfo{author}{\bibfnamefont{F.}~\bibnamefont{Abe}},
  \bibinfo{journal}{Astrophys. J.} \textbf{\bibinfo{volume}{740}},
  \bibinfo{pages}{121} (\bibinfo{year}{2011}).

\bibitem[{\citenamefont{Takahashi and Asada}(2013)}]{Takahashi:2013jqa}
\bibinfo{author}{\bibfnamefont{R.}~\bibnamefont{Takahashi}} \bibnamefont{and}
  \bibinfo{author}{\bibfnamefont{H.}~\bibnamefont{Asada}},
  \bibinfo{journal}{Astrophys J. Lett.} \textbf{\bibinfo{volume}{768}},
  \bibinfo{pages}{L16} (\bibinfo{year}{2013}).

\bibitem[{\citenamefont{Cramer et~al.}(1995)\citenamefont{Cramer, Forward,
  Morris, Visser, Benford, and Landis}}]{Cramer:1994qj}
\bibinfo{author}{\bibfnamefont{J.~G.} \bibnamefont{Cramer}},
  \bibinfo{author}{\bibfnamefont{R.~L.} \bibnamefont{Forward}},
  \bibinfo{author}{\bibfnamefont{M.~S.} \bibnamefont{Morris}},
  \bibinfo{author}{\bibfnamefont{M.}~\bibnamefont{Visser}},
  \bibinfo{author}{\bibfnamefont{G.}~\bibnamefont{Benford}}, \bibnamefont{and}
  \bibinfo{author}{\bibfnamefont{G.~A.} \bibnamefont{Landis}},
  \bibinfo{journal}{Phys. Rev. D} \textbf{\bibinfo{volume}{51}},
  \bibinfo{pages}{3117} (\bibinfo{year}{1995}).

\bibitem[{\citenamefont{Perlick}(2004)}]{Perlick:2003vg}
\bibinfo{author}{\bibfnamefont{V.}~\bibnamefont{Perlick}},
  \bibinfo{journal}{Phys. Rev. D} \textbf{\bibinfo{volume}{69}},
  \bibinfo{pages}{064017} (\bibinfo{year}{2004}).

\bibitem[{\citenamefont{Tsukamoto et~al.}(2012)\citenamefont{Tsukamoto, Harada,
  and Yajima}}]{Tsukamoto:2012xs}
\bibinfo{author}{\bibfnamefont{N.}~\bibnamefont{Tsukamoto}},
  \bibinfo{author}{\bibfnamefont{T.}~\bibnamefont{Harada}}, \bibnamefont{and}
  \bibinfo{author}{\bibfnamefont{K.}~\bibnamefont{Yajima}},
  \bibinfo{journal}{Phys. Rev. D} \textbf{\bibinfo{volume}{86}},
  \bibinfo{pages}{104062} (\bibinfo{year}{2012}).

\bibitem[{\citenamefont{Bambi}(2013)}]{Bambi:2013nla}
\bibinfo{author}{\bibfnamefont{C.}~\bibnamefont{Bambi}},
  \bibinfo{journal}{Phys. Rev. D} \textbf{\bibinfo{volume}{87}},
  \bibinfo{pages}{107501} (\bibinfo{year}{2013}).

\bibitem[{\citenamefont{Zhou et~al.}(2016)\citenamefont{Zhou,
  Cardenas-Avendano, Bambi, Kleihaus, and Kunz}}]{Zhou:2016koy}
\bibinfo{author}{\bibfnamefont{M.}~\bibnamefont{Zhou}},
  \bibinfo{author}{\bibfnamefont{A.}~\bibnamefont{Cardenas-Avendano}},
  \bibinfo{author}{\bibfnamefont{C.}~\bibnamefont{Bambi}},
  \bibinfo{author}{\bibfnamefont{B.}~\bibnamefont{Kleihaus}}, \bibnamefont{and}
  \bibinfo{author}{\bibfnamefont{J.}~\bibnamefont{Kunz}},
  \bibinfo{journal}{Phys. Rev. D} \textbf{\bibinfo{volume}{94}},
  \bibinfo{pages}{024036} (\bibinfo{year}{2016}).

\bibitem[{\citenamefont{Bl{\'a}zquez-Salcedo
  et~al.}(2018)\citenamefont{Bl{\'a}zquez-Salcedo, Chew, and
  Kunz}}]{Blazquez-Salcedo:2018ipc}
\bibinfo{author}{\bibfnamefont{J.~L.} \bibnamefont{Bl{\'a}zquez-Salcedo}},
  \bibinfo{author}{\bibfnamefont{X.~Y.} \bibnamefont{Chew}}, \bibnamefont{and}
  \bibinfo{author}{\bibfnamefont{J.}~\bibnamefont{Kunz}},
  \bibinfo{journal}{Phys. Rev. D} \textbf{\bibinfo{volume}{98}},
  \bibinfo{pages}{044035} (\bibinfo{year}{2018}).

\bibitem[{\citenamefont{Hajicek}(1983{\natexlab{a}})}]{Hajicek_1983a}
\bibinfo{author}{\bibfnamefont{P.}~\bibnamefont{Hajicek}},
  \bibinfo{journal}{Proceedings of the Royal Society of London. Series A,
  Mathematical and Physical Sciences} \textbf{\bibinfo{volume}{386}},
  \bibinfo{pages}{223} (\bibinfo{year}{1983}{\natexlab{a}}), ISSN
  \bibinfo{issn}{00804630}.

\bibitem[{\citenamefont{Hajicek}(1983{\natexlab{b}})}]{Hajicek_1983}
\bibinfo{author}{\bibfnamefont{P.}~\bibnamefont{Hajicek}},
  \bibinfo{journal}{Journal of Physics A: Mathematical and General}
  \textbf{\bibinfo{volume}{16}}, \bibinfo{pages}{1191}
  (\bibinfo{year}{1983}{\natexlab{b}}).

\bibitem[{\citenamefont{{Degen}}(1987)}]{1987GReGr..19..739D}
\bibinfo{author}{\bibfnamefont{F.}~\bibnamefont{{Degen}}},
  \bibinfo{journal}{General Relativity and Gravitation}
  \textbf{\bibinfo{volume}{19}}, \bibinfo{pages}{739} (\bibinfo{year}{1987}).

\bibitem[{\citenamefont{Dzhunushaliev
  et~al.}(2014{\natexlab{a}})\citenamefont{Dzhunushaliev, Folomeev, Hoffmann,
  Kleihaus, and Kunz}}]{Dzhunushaliev:2014bya}
\bibinfo{author}{\bibfnamefont{V.}~\bibnamefont{Dzhunushaliev}},
  \bibinfo{author}{\bibfnamefont{V.}~\bibnamefont{Folomeev}},
  \bibinfo{author}{\bibfnamefont{C.}~\bibnamefont{Hoffmann}},
  \bibinfo{author}{\bibfnamefont{B.}~\bibnamefont{Kleihaus}}, \bibnamefont{and}
  \bibinfo{author}{\bibfnamefont{J.}~\bibnamefont{Kunz}},
  \bibinfo{journal}{Phys. Rev. D} \textbf{\bibinfo{volume}{90}},
  \bibinfo{pages}{124038} (\bibinfo{year}{2014}{\natexlab{a}}).

\bibitem[{\citenamefont{Hoffmann et~al.}(2017)\citenamefont{Hoffmann,
  Ioannidou, Kahlen, Kleihaus, and Kunz}}]{Hoffmann:2017jfs}
\bibinfo{author}{\bibfnamefont{C.}~\bibnamefont{Hoffmann}},
  \bibinfo{author}{\bibfnamefont{T.}~\bibnamefont{Ioannidou}},
  \bibinfo{author}{\bibfnamefont{S.}~\bibnamefont{Kahlen}},
  \bibinfo{author}{\bibfnamefont{B.}~\bibnamefont{Kleihaus}}, \bibnamefont{and}
  \bibinfo{author}{\bibfnamefont{J.}~\bibnamefont{Kunz}},
  \bibinfo{journal}{Phys. Rev. D} \textbf{\bibinfo{volume}{95}},
  \bibinfo{pages}{084010} (\bibinfo{year}{2017}).

\bibitem[{\citenamefont{Corichi et~al.}(2000)\citenamefont{Corichi, Nucamendi,
  and Sudarsky}}]{Corichi:2000dm}
\bibinfo{author}{\bibfnamefont{A.}~\bibnamefont{Corichi}},
  \bibinfo{author}{\bibfnamefont{U.}~\bibnamefont{Nucamendi}},
  \bibnamefont{and} \bibinfo{author}{\bibfnamefont{D.}~\bibnamefont{Sudarsky}},
  \bibinfo{journal}{Phys. Rev. D} \textbf{\bibinfo{volume}{62}},
  \bibinfo{pages}{044046} (\bibinfo{year}{2000}), \eprint{gr-qc/0002078}.

\bibitem[{\citenamefont{Ashtekar et~al.}(2001)\citenamefont{Ashtekar, Corichi,
  and Sudarsky}}]{Ashtekar:2000nx}
\bibinfo{author}{\bibfnamefont{A.}~\bibnamefont{Ashtekar}},
  \bibinfo{author}{\bibfnamefont{A.}~\bibnamefont{Corichi}}, \bibnamefont{and}
  \bibinfo{author}{\bibfnamefont{D.}~\bibnamefont{Sudarsky}},
  \bibinfo{journal}{Class. Quant. Grav.} \textbf{\bibinfo{volume}{18}},
  \bibinfo{pages}{919} (\bibinfo{year}{2001}), \eprint{gr-qc/0011081}.

\bibitem[{\citenamefont{Ascher et~al.}(1979)\citenamefont{Ascher, Christiansen,
  and Russell}}]{colsys}
\bibinfo{author}{\bibfnamefont{U.}~\bibnamefont{Ascher}},
  \bibinfo{author}{\bibfnamefont{J.}~\bibnamefont{Christiansen}},
  \bibnamefont{and} \bibinfo{author}{\bibfnamefont{R.~D.}
  \bibnamefont{Russell}}, \bibinfo{journal}{Math. Comput.}
  \textbf{\bibinfo{volume}{33}}, \bibinfo{pages}{659} (\bibinfo{year}{1979}).

\bibitem[{\citenamefont{Kleihaus et~al.}(2020)\citenamefont{Kleihaus, Kunz, and
  Kanti}}]{Kleihaus:2020qwo}
\bibinfo{author}{\bibfnamefont{B.}~\bibnamefont{Kleihaus}},
  \bibinfo{author}{\bibfnamefont{J.}~\bibnamefont{Kunz}}, \bibnamefont{and}
  \bibinfo{author}{\bibfnamefont{P.}~\bibnamefont{Kanti}},
  \bibinfo{journal}{Phys. Rev. D} \textbf{\bibinfo{volume}{102}},
  \bibinfo{pages}{024070} (\bibinfo{year}{2020}), \eprint{2005.07650}.

\bibitem[{\citenamefont{Chew and Lim}(\noop{2020}in the
  preparation)}]{soonHiggsfinite}
\bibinfo{author}{\bibfnamefont{X.~Y.} \bibnamefont{Chew}} \bibnamefont{and}
  \bibinfo{author}{\bibfnamefont{K.-G.} \bibnamefont{Lim}}
  (\bibinfo{year}{\noop{2020}in the preparation}).

\bibitem[{\citenamefont{Dzhunushaliev
  et~al.}(2013{\natexlab{b}})\citenamefont{Dzhunushaliev, Folomeev, Kleihaus,
  and Kunz}}]{Dzhunushaliev:2013lna}
\bibinfo{author}{\bibfnamefont{V.}~\bibnamefont{Dzhunushaliev}},
  \bibinfo{author}{\bibfnamefont{V.}~\bibnamefont{Folomeev}},
  \bibinfo{author}{\bibfnamefont{B.}~\bibnamefont{Kleihaus}}, \bibnamefont{and}
  \bibinfo{author}{\bibfnamefont{J.}~\bibnamefont{Kunz}},
  \bibinfo{journal}{Phys. Rev. D} \textbf{\bibinfo{volume}{87}},
  \bibinfo{pages}{104036} (\bibinfo{year}{2013}{\natexlab{b}}).

\bibitem[{\citenamefont{Dzhunushaliev
  et~al.}(2014{\natexlab{b}})\citenamefont{Dzhunushaliev, Folomeev, Kleihaus,
  and Kunz}}]{Dzhunushaliev:2014mza}
\bibinfo{author}{\bibfnamefont{V.}~\bibnamefont{Dzhunushaliev}},
  \bibinfo{author}{\bibfnamefont{V.}~\bibnamefont{Folomeev}},
  \bibinfo{author}{\bibfnamefont{B.}~\bibnamefont{Kleihaus}}, \bibnamefont{and}
  \bibinfo{author}{\bibfnamefont{J.}~\bibnamefont{Kunz}},
  \bibinfo{journal}{Phys. Rev. D} \textbf{\bibinfo{volume}{89}},
  \bibinfo{pages}{084018} (\bibinfo{year}{2014}{\natexlab{b}}).

\bibitem[{\citenamefont{Aringazin et~al.}(2015)\citenamefont{Aringazin,
  Dzhunushaliev, Folomeev, Kleihaus, and Kunz}}]{Aringazin:2014rva}
\bibinfo{author}{\bibfnamefont{A.}~\bibnamefont{Aringazin}},
  \bibinfo{author}{\bibfnamefont{V.}~\bibnamefont{Dzhunushaliev}},
  \bibinfo{author}{\bibfnamefont{V.}~\bibnamefont{Folomeev}},
  \bibinfo{author}{\bibfnamefont{B.}~\bibnamefont{Kleihaus}}, \bibnamefont{and}
  \bibinfo{author}{\bibfnamefont{J.}~\bibnamefont{Kunz}}, \bibinfo{journal}{J.
  Cosmol. Astropart. Phys.} \textbf{\bibinfo{volume}{2015}},
  \bibinfo{pages}{005} (\bibinfo{year}{2015}).

\bibitem[{\citenamefont{Greene et~al.}(1993)\citenamefont{Greene, Mathur, and
  O'neill}}]{greene1993eluding}
\bibinfo{author}{\bibfnamefont{B.~R.} \bibnamefont{Greene}},
  \bibinfo{author}{\bibfnamefont{S.~D.} \bibnamefont{Mathur}},
  \bibnamefont{and} \bibinfo{author}{\bibfnamefont{C.~M.}
  \bibnamefont{O'neill}}, \bibinfo{journal}{Phys. Rev. D}
  \textbf{\bibinfo{volume}{47}}, \bibinfo{pages}{2242} (\bibinfo{year}{1993}).

\bibitem[{\citenamefont{Winstanley and
  Mavromatos}(1995)}]{winstanley1995instability}
\bibinfo{author}{\bibfnamefont{E.}~\bibnamefont{Winstanley}} \bibnamefont{and}
  \bibinfo{author}{\bibfnamefont{N.~E.} \bibnamefont{Mavromatos}},
  \bibinfo{journal}{Phys. Lett. B} \textbf{\bibinfo{volume}{352}},
  \bibinfo{pages}{242} (\bibinfo{year}{1995}).

\bibitem[{\citenamefont{Mavromatos and
  Winstanley}(1996)}]{mavromatos1996aspects}
\bibinfo{author}{\bibfnamefont{N.~E.} \bibnamefont{Mavromatos}}
  \bibnamefont{and}
  \bibinfo{author}{\bibfnamefont{E.}~\bibnamefont{Winstanley}},
  \bibinfo{journal}{Phys. Rev. D} \textbf{\bibinfo{volume}{53}},
  \bibinfo{pages}{3190} (\bibinfo{year}{1996}).

\end{thebibliography}

\end{document}